\documentclass[12pt]{article}
\pdfoutput=1
\usepackage{jcapmod}
\usepackage{booktabs}
\usepackage{mathtools}
\usepackage[english]{babel}
\usepackage{amsmath,amssymb,amsbsy,amstext, amsthm}
\usepackage[usenames,dvipsnames]{xcolor}
\usepackage{hyperref}
\hypersetup{
    urlcolor=NavyBlue,
    citecolor=NavyBlue,
    linkcolor=NavyBlue,
}%
\usepackage{array}
\usepackage{url} 
\usepackage{slashed}
\usepackage{multicol}
\usepackage{blkarray}
\usepackage{enumerate}
\usepackage{graphicx}
\usepackage{amsfonts}
\usepackage{amssymb}
  \usepackage{empheq}
  \usepackage{tcolorbox}
\usepackage{xcolor}
\usepackage{soul}
\usepackage{colortbl}

\usepackage{float}
\usepackage[utf8]{inputenc}
\RequirePackage{color}
\usepackage[normalem]{ulem}
\graphicspath{{plots/}}

\newcommand{\ml}[1]{{\ensuremath{\,\scalebox{0.8}{(#1)}}}}

\usepackage{colortbl}
\definecolor{green2}{cmyk}{0, 1, 0.5, 0}
\definecolor{lightgreen}{cmyk}{0.2, 0, 0.2, 0.2}
\definecolor{lightgray}{cmyk}{0.1,0.2,0,0.1}
\definecolor{lightgray2}{cmyk}{0.4,0.4,0,0.8}
\definecolor{black}{cmyk}{1.0,1.0,1.0,1.0}

\allowdisplaybreaks[1]

\usepackage{colortbl}
\definecolor{lightgreen}{cmyk}{0.2, 0, 0.2, 0.2}
\definecolor{lightgray}{cmyk}{0.1,0.2,0,0.1}
\definecolor{lightgray2}{cmyk}{0.1,0.1,0,0.1}

\makeatletter
\newlength{\apb@width}
\newcommand{\autoparbox}[2][c]{\settowidth{\apb@width}{#2}\parbox[#1]{\apb@width}{#2}}

\makeatother

\numberwithin{equation}{section}

\def\be{\begin{equation}}
\def\ee{\end{equation}}

\def\bea{\begin{eqnarray}}
\def\eea{\end{eqnarray}}

\def\Mp{M_{\rm Pl}}

\def\0{{\boldsymbol 0}}

\usepackage{setspace} 
\begin{document}

\begin{titlepage}

\setcounter{page}{1} \baselineskip=15.5pt \thispagestyle{empty}

\bigskip\

\vspace{1cm}
\begin{center}

{\fontsize{20}{28}\selectfont  \sffamily \bfseries {Bounded Dark Energy}}

\end{center}

\vspace{-0.1cm}

\begin{center} 
{\fontsize{13}{30}
 Giulia Borghetto$^{a,\!}$~\footnote{\texttt{giulia.borghetto@gmail.com}}, Ameek Malhotra$^{a,\!}$~\footnote{\texttt{ameek.malhotra@swansea.ac.uk}}, Gianmassimo Tasinato$^{a,b\!}$~\footnote{\texttt{g.tasinato2208@gmail.com}}, Ivonne Zavala$^{a,\!}$~\footnote{\texttt{e.i.zavalacarrasco@swansea.ac.uk}}
} 
\end{center}

\begin{center}

\vskip 6pt
\textsl{$^a$ Physics Department, Swansea University, SA2 8PP, UK}
\\
\textsl{$^{b}$ Dipartimento di Fisica e Astronomia, Universit\`a di Bologna,\\
 INFN, Sezione di Bologna, I.S. FLAG, viale B. Pichat 6/2, 40127 Bologna,   Italy}
\vskip 4pt

\end{center}

\vspace{1.cm}
\hrule \vspace{0.3cm}
\noindent
Recent cosmological observations suggest that the dark energy equation of state may have  changed in the latter stages of cosmic history. We introduce a quintessence scenario, termed bounded dark energy, capable of explaining this feature in a technically natural way. Our approach is motivated from a bottom-up perspective, based on the concept of mirage cut-off, where we demonstrate the stability of the quintessence potential against large quantum corrections. At the same time, the bounded dark energy framework aligns well with top-down considerations motivated from quantum gravity arguments. We employ both human-driven insights and machine learning techniques to identify explicit realizations of bounded dark energy models.
We then perform an analysis based on Markov Chain Monte-Carlo to assess their predictions against CMB, galaxy surveys, and supernova data,  showing that bounded dark energy provides a good fit to current observations. We also discuss how upcoming measurements can further test and refine our proposal.
\vskip 10pt
\hrule
 
\vspace{0.4cm}
 \end{titlepage}


\section{Motivations}
\label{sec_mot}

\begin{figure}[H]
    \centering
    \includegraphics[width=0.7\linewidth]{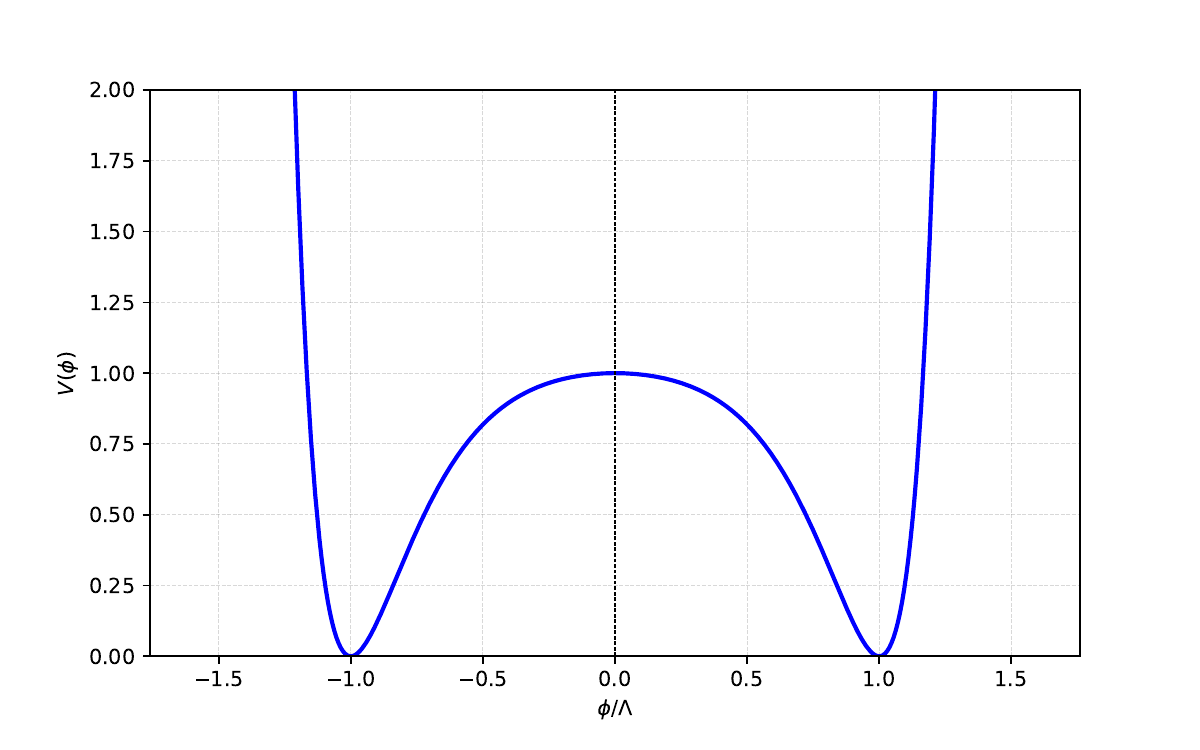}
    \caption{\small \it The typical shape of a quintessence potential describing bounded dark energy scenarios.}
    \label{fig_typ} 
\end{figure}

Understanding the physics of dark energy remains one of the most pressing open problems in modern physics \cite{Weinberg:1988cp}. Numerous questions arise concerning the dark sector, both from observational and theoretical   perspectives. Is dark energy dynamical? Why is its magnitude so small compared to theoretical expectations? Is there any mechanism to ensure  stability of its properties against quantum corrections?
We address 
some of these fundamental questions
 investigating quintessential potentials characterized by a scalar profile of the type depicted in Figure \ref{fig_typ}.

\noindent
The quintessential scalar starts
rolling from the local maximum of the potential, and its trajectory is bounded in field space
by the steep growth of $V(\phi)$. 
Our reasons are the following:
\begin{enumerate}
\item Recent cosmological results \cite{DES:2024tys, DESI:2024mwx,DESI:2024hhd, DES:2025bxy} 
 suggest 
that dark energy (DE) is dynamical.\footnote{The significance of these results, possibility of systematic effects and other different interpretations have been extensively discussed in~\cite{Colgain:2024xqj,Carloni:2024zpl,Park:2024jns,Wang:2024rjd,Cortes:2024lgw,Wang:2024pui,Dinda:2024kjf,Croker:2024jfg,Wang:2024hks,Luongo:2024fww,Mukherjee:2024ryz,Wang:2024dka,Efstathiou:2024xcq,Tada:2024znt,Yin:2024hba,Berghaus:2024kra,Shlivko:2024llw,Alestas:2024eic,Sohail:2024oki,RoyChoudhury:2024wri,Giare:2024gpk,Giare:2024smz,Giare:2024ocw,Giare:2025pzu,Keeley:2025stf,DES:2025tir,Chan-GyungPark:2025cri,Taylor:2024whh,Colgain:2024mtg,Payeur:2024dnq,Berbig:2024aee,Sapone:2024ltl,Gao:2024ily,Tamayo:2025wiy,Ormondroyd:2025exu}.}  Its  behavior is well described 
by the so-called CPL parameterisation  \cite{Chevallier:2000qy, Linder:2002et}, with a DE equation
of state scaling as $w(z)\,=\,w_0+w_a\,z/(1+z)$. 
In fact, analysis from the first year of observations from recent Baryon Acoustic Oscillation (BAO) surveys as well as recent Supernovae datasets reveal a marked preference for the CPL parameterization over 
 LCDM~\cite{DESI:2024mwx,DESI:2024hhd}.  
 A dynamical DE 
following the CPL prescription crosses the phantom divide at $w=-1$, thus  it is  important to design theoretically consistent models  avoiding such pathological
behavior. Quintessence scenarios \cite{Ratra:1987rm,Caldwell:1997ii}, by construction, ensure that $w\ge-1$. 
Among this broad class of models, thawing quintessence  \cite{Frieman:1995pm} describes a setup where the scalar field begins to acquire significant kinetic energy relatively late during cosmic evolution. Such behavior  typically occurs when the scalar initially evolves from a flat region of its potential (for example around a local maximum), which then steepens, leading to an increase in the scalar kinetic energy (see  Fig.~\ref{fig_typ}). The dynamics naturally lead to the condition $w_a < 0$ in the context of the CPL parameterization~\cite{Caldwell:2005tm}. Consequently, thawing models are particularly relevant in light of current observational data.
  A notable example~\footnote{Besides hilltop,
 other scalar potentials can fit current data well. See e.g.~\cite{Bhattacharya:2024hep, Bhattacharya:2024kxp} for recent cosmological analysis of string motivated  quintessence models.} is the class of hilltop quintessence models \cite{Dutta:2008qn},
for example recently studied in \cite{Bhattacharya:2024kxp} in the context of string-motivated quintessence scenarios.

Moreover,
quintessence potentials with  sudden changes in  their profile can effectively replicate  accurately the dynamics of dark energy as described by the CPL parameterization, and at the same time avoid the phantom crossing divide: see e.g.
\cite{Shlivko:2024llw,Tada:2024znt,Gialamas:2024lyw,Notari:2024rti}.
This leads us to explore models, as shown in Fig.~\ref{fig_typ}, that exhibit particularly steep declines, following a flat region near the potential origin.

\item  
Thawing quintessence, based on a scalar field framework and hinted at by recent observations, introduces an additional layer of theoretical complexity for DE models. Not only must we explain why the DE scale is so small compared to other energy scales in particle physics \cite{Weinberg:1988cp}, but we must also identify mechanisms to stabilize the distinctive shape of the scalar potential against quantum loop corrections. We aim to link the two problems and address them using the concept of  {\it mirage cut-off} introduced in  \cite{Cheung:2024wme}.

\smallskip

 The essential idea is that the scalar potential is constituted by a  large number of contributions in its Taylor expansion, weighted by an overall small quantity:
 \be
 \label{eq_inans}
 V(\phi)\,=\,\epsilon\,\Lambda^4\,\sum_{n=0}^{\infty}
 \,\frac{c_n}{n!}\,\left(\frac{\phi^2}{\Lambda^2} \right)^n
 \ee
 with $\Lambda$
 an energy scale, $\epsilon$ a very small number, and $c_n$
 order-one coefficients. 
 In our context, $\Lambda$ represents the model cut-off, and $V_0 = \epsilon \Lambda^4$ sets the dark energy (DE) scale, which 
 by having  a small $\epsilon$ 
 is significantly suppressed with respect to the cut-off
 scale $\Lambda^4$ (a small DE scale is of course  well
 motivated by  observations). Thanks to the smallness of $\epsilon$, the flat region of the potential remains {\it stable under quantum corrections} \cite{Cheung:2024wme}. 
   We explore these ideas  in Section~\ref{sec_bu}, applying them to DE, and discussing how generic potentials like \eqref{eq_inans} naturally lead to boundaries in field space excursions, thanks to the (unavoidable) steep rises in their magnitude at finite values of the scalar field. Besides
  developing simple representative models in Section \ref{sec_bu}, in Section~\ref{sec_ml} 
  we  use  machine learning techniques to find
 realisations of bounded DE setup with the correct features to accurately fit 
current cosmological data.

From a top-down perspective, a sudden change in the potential and the emergence of boundaries in field space may signal the presence of topological objects, as predicted by the cobordism conjecture\footnote{The cobordism conjecture suggests that in any consistent theory of quantum gravity all allowed configurations must not carry any cobordism charge, implying that the cobordism group must be trivial (i.e., vanish).} \cite{McNamara:2019rup, Montero:2020icj,Blumenhagen:2023abk}.
These could manifest as interpolating domain walls between consistent, cobordant theories (or vacua) at a finite distance in both field space and spacetime, as recently discussed in \cite{Buratti:2021fiv}.

\item  Potentials as Fig.~\ref{fig_typ}, resembling  a Mexican hat-like shape,
naturally find realisations in symmetry breaking field theory scenarios. When coupling the DE scalar to matter fields, such
   symmetry breaking pattern  also provides a natural screening mechanism for fifth forces, by means of  the symmetron mechanism of \cite{Hinterbichler:2010es}, while simultaneously suggesting intriguing possibilities for setting the scalar field's initial conditions.  
We  outline these arguments  in Section~\ref{sec_sb}.
\end{enumerate}
The previous theoretical ideas
are realised in explicit bounded dark energy scenarios, that we compare against the most recent cosmological data in Section \ref{sec_costest}.

\section{Theoretical model building}
\label{sec_th}

A dynamical dark energy, if supported by data, requires model building efforts for
establishing  setups which are theoretically under control. 
We   consider hilltop quintessential potentials with sharp boundaries
 in field space, as the one pictorially represented in Fig.~\ref{fig_typ}.
  We provide
 arguments in favor of the technical
 stability of such potentials under quantum 
 corrections,  
  building on \cite{Cheung:2024wme}.

\subsection{A bottom-up perspective}
\label{sec_bu}
We  start by progressively developing arguments supporting the stability of potentials like those shown in Fig.~\ref{fig_typ} under loop corrections, using 
an effective field theory perspective.
We  link the reduced size of quantum loop corrections with the smallness of the DE scale.
The essential
ingredient we use is the concept of mirage cut-off  \cite{Cheung:2024wme} applied to DE scenarios.

\smallskip

 The scalar potentials $V(\phi)$ we consider exhibit a relatively flat plateau with a height $V_0$ around the origin $\phi \sim 0$. This is followed by a steep decline in the potential amplitude at a finite value of the scalar field, which then transitions into a rapid increase and reaches values much greater than $V_0$. 
 Examples are quintessence hilltop scenarios
\cite{Dutta:2008qn}, being equipped by 
  a local maximum 
 around which the scalar  starts slowly rolling with an almost constant energy density -- mimicking a cosmological constant at early times. Then, the flat region nearby the maximum changes slope and becomes steep, making the scalar roll faster from a certain position onwards.
 A well-motivated, simple  hilltop model
 is the quartic Higgs potential 
 \be 
 \label{eq_qpot}
 V(\phi)\,=\,V_0 \left(1-\frac{\phi^2}{\Lambda^2}\right)^2 \,,
 \ee
 tested for example in \cite{Bhattacharya:2024kxp}   against the most recent observational results. 
 We proceed to analyse  some of its possible generalizations.
 
\begin{figure}[t!]
    \centering
    \includegraphics[width=0.49\linewidth]{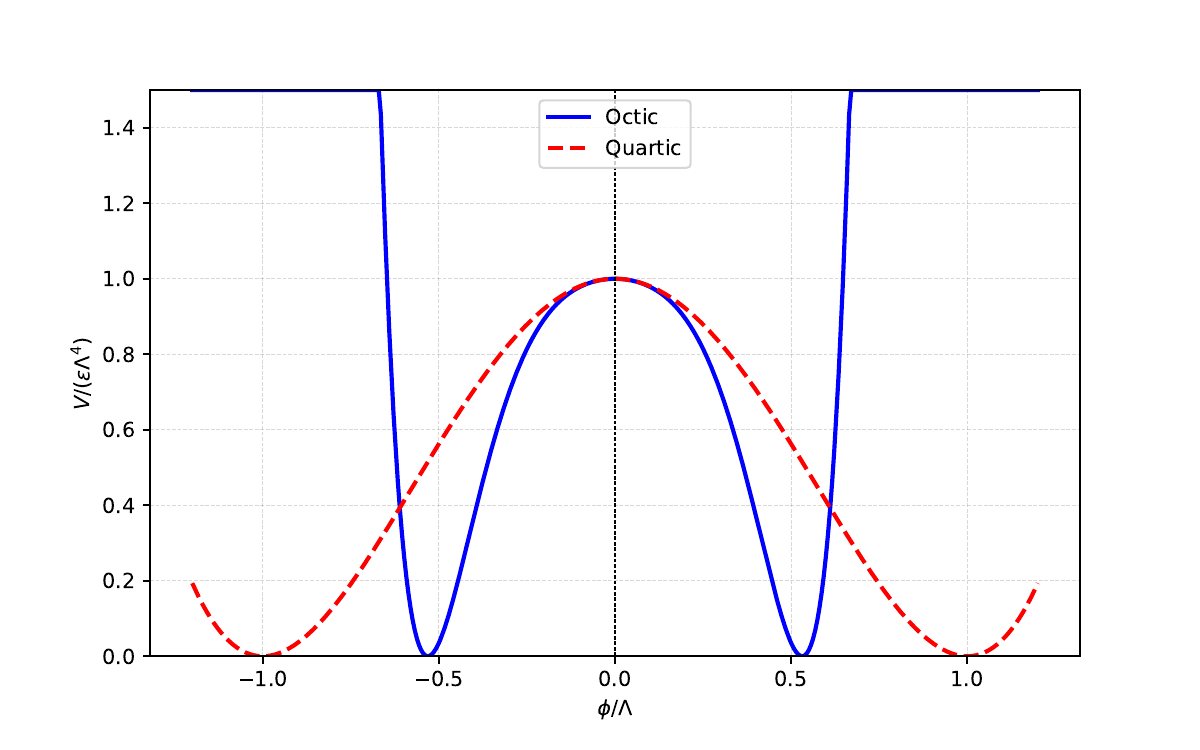}
    \vspace{0.5cm}
    \includegraphics[width=0.49
    \linewidth]{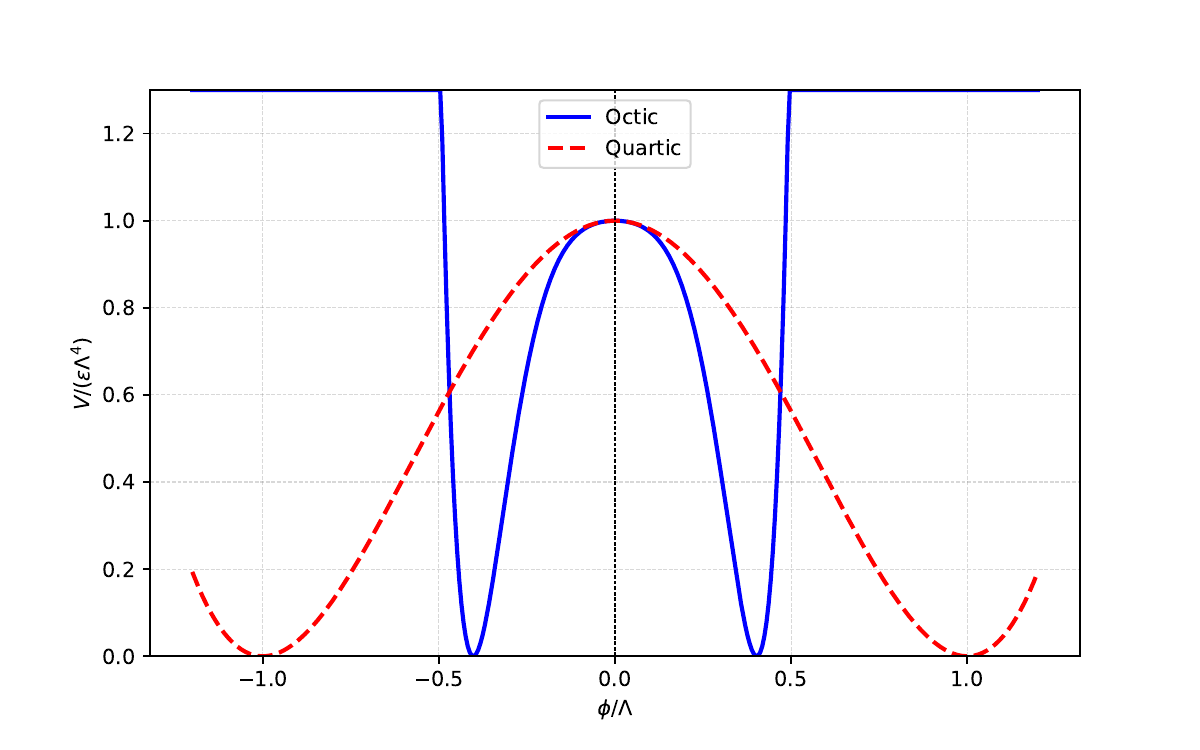}
    \caption{\it  Representation of the hilltop octic potential \eqref{eq_oct} (blue line), versus quartic potential  ($c_0=0$, red dashed line). {\bf Left} $c_0=36$ {\bf Right} $c_0=128$. Larger values
    of $c_0$ make the potential descent steeper and steeper.}
    \label{fig_pot1}
\end{figure}

\subsubsection*{The hilltop octic potential}
 We  first consider a one-parameter
 extension of eq  \eqref{eq_qpot}  examining the
 potential
 \be
 \label{eq_oct}
 V_{\rm oct}(\phi)\,=\,\epsilon\,\Lambda^4 \left(1-\frac{\phi^2}{\Lambda^2} 
 -\frac{c_0}{4}\,\frac{\phi^4}{\Lambda^4} \right)^2\,.
 \ee
 We dub this potential {\it octic} since when expanding the parenthesis the scalar acquires powers up to level
eight. In the limit $c_0\to 0$
we recover the structure of quartic potential. The octic potential has a maximum at $\phi=0$.
The dimensionful quantity $\Lambda$, together with $c_0$, control 
the size of the flat region around the maximum of the potential. The value 
of the potential at the maximum is
given  by $V_0= \epsilon \Lambda^4$. For our purposes of applying the system
to DE, we consider very small values for
 $\epsilon$, which enters  as an overall factor of our potential.
 A quintessence model associated with
  potential \eqref{eq_oct} is not too sophisticated and it has not yet
  all the theoretically appealing features 
of the potentials we will consider next. Nevertheless it has the following interesting properties:  

 \begin{itemize}
\item[-] First, the parameter \( c_0 \) can   steepen the descending well of the potential compared to the quartic case, resulting in a more abrupt transition to the dynamical dark energy regime—see Fig.~\ref{fig_pot1}. The potential minimum 
is followed by a steep rise associated
with the large power \( \phi^8 \).
Sharp features and sudden transitions could have intriguing implications when confronted with recent findings in cosmological data sets.
 We develop this
topic in Section \ref{sec_costest}. 
\item[-] Second,  potential \eqref{eq_oct} gives
a first glance on the phenomenon of
{\it mirage cut-off} \cite{Cheung:2024wme}. Expanding \eqref{eq_oct}, we observe that in the limit $\epsilon \ll 1$, the apparent cut-offs at levels six and eight in the expansion, $\epsilon^{-1/2}\,\Lambda$ and $\epsilon^{-1/4}\,\Lambda$, can be parametrically  larger than the  cut-off of the theory,   of order $\Lambda$. If we were continuing the series with richer potentials, the size of the mirage cut-offs would become smaller and smaller, and approach $\Lambda$ more and more.  
Such behavior can have interesting
implications for the stability of the set-up under quantum corrections  \cite{Cheung:2024wme}, as
we are going to discuss next.
\end{itemize}

In fact,
we now generalize the idea behind eq \eqref{eq_oct} and we 
pass to consider quintessence hilltop potentials made by an {\it infinite
series} of powers of $\phi^2/\Lambda^2$ terms, as
\be
\label{eq_tay}
 V(\phi)\,=\,\epsilon\,\Lambda^4\,\sum_{n=0}^{\infty}
 \,\frac{c_n}{n!}\,\left(\frac{\phi^2}{\Lambda^2} \right)^n\,,
 \ee
 with $c_n$ dimensionless
  coefficients  of order one, and the overall coefficient
 $\epsilon$ a very small
  number, making the overall scale $V_0 = \epsilon \Lambda^4 \ll \Lambda^4$.
  The expansion \eqref{eq_tay} exhibits the phenomenon of 
mirage cut-off to which we alluded above \cite{Cheung:2024wme}. The various terms
exhibit cut-off scales scaling progressively as $\epsilon^{-1/2}\,\Lambda$, $\epsilon^{-1/4}\,\Lambda$, etc. Only probing the theory  at high multiplicities of order $n > \ln{\epsilon^{-1/2}}$ the true cut-off of the theory, $\Lambda$,  becomes manifest \cite{Cheung:2024wme}. Hence we  have  the following chain of inequalities, controlled by the small $\epsilon$:
\be
\Lambda_{\rm DE}\,=\,\epsilon^{1/4}\,\Lambda
\,
\ll
\,
\Lambda 
\,
\ll
\,\epsilon^{-1/2}\,\Lambda\,=\,\Lambda_{\rm mirage}
\label{eq_chain}
\ee
where on the left is the DE scale (which we wish to be small). On the right the first mirage cut-off scale. In the middle, the
true cut-off scale $\Lambda$ of the theory, which can be probed at high multiplicities~\footnote{In Section \ref{sec_td} we suggest ideas to realise scenarios with  a small overall factor
$\epsilon$ in front of the potential \eqref{eq_tay}, by considering a geometrical, higher dimensional perspective to DE model building.}.
Inequality \eqref{eq_chain} makes clear that the smallness of $\epsilon$ is both related with the smallness of the DE scale and with the concept of mirage cut-off.

A peculiar feature of eq \eqref{eq_tay} is that
it generally leads to pronounced features in the potential \cite{Cheung:2024wme}. Since the coefficients \( c_n \) are of order one by hypothesis, the Taylor series converges, and defines an entire function in the complex plane. By Picard's Little Theorem, a non-constant entire function must assume all complex values, except possibly one. When restricted to the real domain, this feature implies that \( V(\phi) \) will attain values much larger than \( V_0 \) for a finite \( \bar{\phi}  \) --  at least of order \( \Lambda \) -- effectively bounding the possible field excursions in a dynamical scalar setting.
For the same reason, depending on the model and the signs of \( c_n \), the potential \( V(\phi) \) may also take values much smaller than \( V_0 \) for \( \phi < \bar{\phi} \), possibly leading to local minima. See e.g. Fig.~\ref{fig_typ} for a possible realization.

\subsubsection*{The hilltop exponential  potential}
\begin{figure}[t!]
    \centering
    \includegraphics[width=0.49\linewidth]{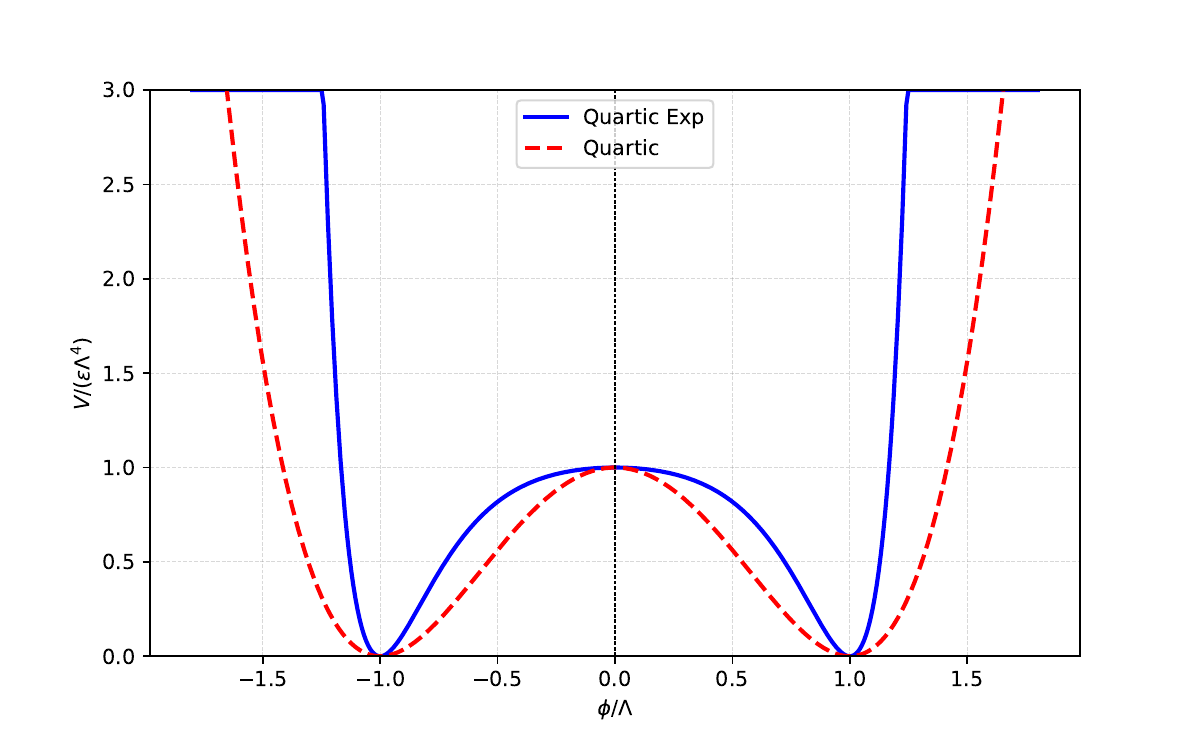}
    \vspace{0.5cm}
    \includegraphics[width=0.49\linewidth]{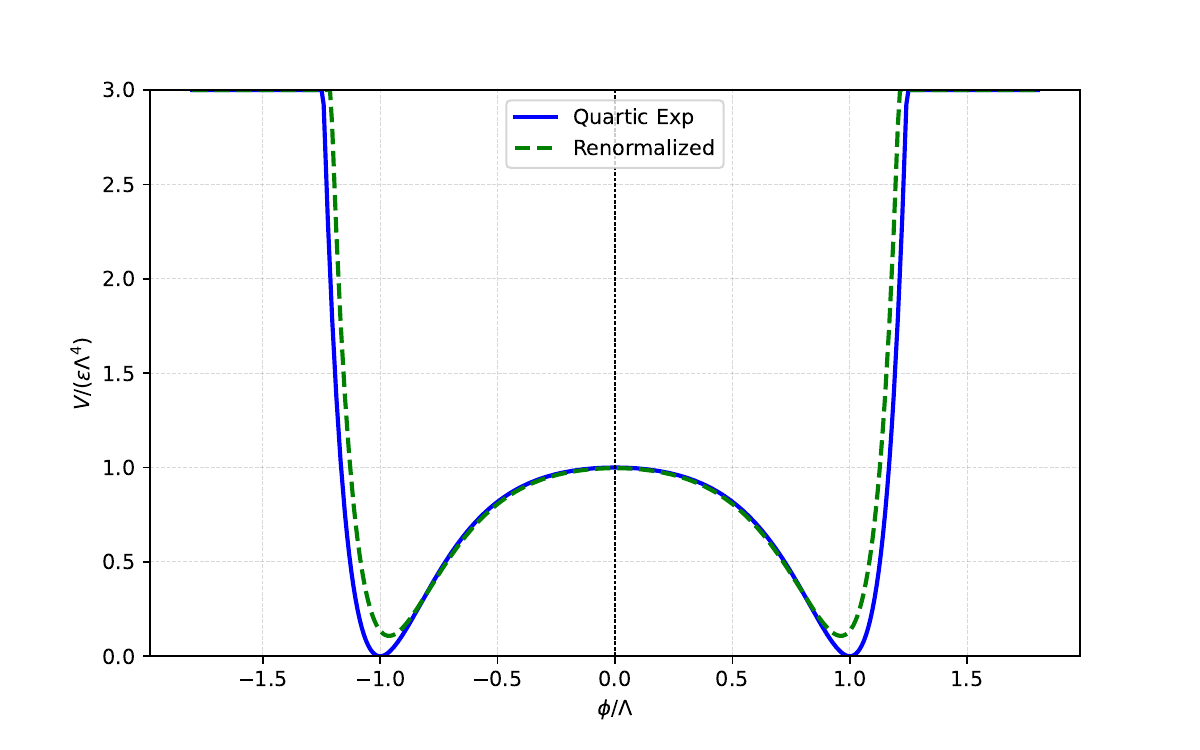}
    \caption{\it {\bf Left panel}: Representation
    of the hilltop  exponential 
    potential \eqref{exphila}  against the quadratic case (blue $\lambda=1.5$, dashed red $\lambda=0$).
     {\bf Right panel}: Hilltop exponential   
    potential \eqref{exphila}  (blue, $\lambda=1.5$)  against  its renormalized version \eqref{eq_renp} (dashed green). }
    \label{fig_pot2}
\end{figure}

  In order to be concrete, we
  consider examples where the infinite
  series \eqref{eq_tay} converges to known functions.
We  
  focus
  on a hilltop exponential potential~\footnote{This potential profile is based on a heuristic human guess. However, as we will learn in   section \ref{sec_ml}, machine learning can achieve better results when  theoretical
  arguments are 
  applied to real data.}
  \be
V(\phi)\,=\,\epsilon\,\Lambda^4\,\left(1-
 \frac{\phi^2}{\Lambda^2}
\right)^2\,e^{\frac{\lambda\,\phi^2}{\Lambda^2}}
\label{exphila}
  \ee
  with the characteristics described above.
Once expanded in a Taylor series, the potential takes the form \eqref{eq_tay}. The dimensionless parameter $\lambda$ in the exponent quantifies deviations from the quartic potential (with \( \lambda =0 \)). The small parameter $\epsilon$ controls the DE  scale, see eq \eqref{eq_chain}. The exponential correction to the quartic case introduces steep features in the potential, as illustrated in Fig.~\ref{fig_pot2} (left panel), while also flattening it near the origin. 
The second derivative of the potential at $\phi = 0$ satisfies 
\be\label{eq:VppVExp}
\frac{V''}{V} \bigg|_{\phi=0} = -\frac{2(2-\lambda)}{\Lambda^2}.
\ee
For $\lambda < 2$  -- condition that we assume from now on -- the potential has minima at $\phi = \pm \Lambda$, as in the quartic case. Beyond these minima, the potential grows steeply, halting any further field excursion, and giving rise to  {\it bounded dark energy} scenario.

\subsubsection*{Stability of bounded DE potentials under
loop corrections}

From a theoretical perspective, an interesting
property of the bounded DE model 
 of eq \eqref{exphila} is that the potential
 shape is technically
 stable  under 
 quantum corrections, as shown in \cite{Cheung:2024wme}. The idea exploits
 the smallness of the parameter $\epsilon$,
which is of course well motivated in a DE setup which requires
a tiny value for the DE scale -- see eq \eqref{eq_chain}. The work \cite{Cheung:2024wme} develops a regularization scheme for power-law divergences
based on the true cut-off $\Lambda$ of the theory as regularizing scale (which
is much smaller than the mirage cut-off scale). 
  Including power-law divergences only,  we can work at linear
order in the small parameter $\epsilon$ and include loop diagrams with a single insertion of the potential:   \cite{Cheung:2024wme} shows that 
the resulting renormalized potential  at all loops is
\bea
V_{\rm ren}(\phi)&=&\frac{\sqrt{8\pi}}{\Lambda}
\int_{-\infty}^{+\infty}
\,d \tilde \phi\,e^{-\frac{8 \pi^2(\tilde \phi-\phi)^2}{\Lambda^2}}\,V(\tilde \phi)\,,
\label{eq_renpa}
\\
&=&\frac{\sqrt{2} \pi\,\epsilon
\,\Lambda^4}{(8 \pi^2-\lambda)^{5/2}}\,W(\phi)
e^{\frac{
8 \pi^2 \lambda \phi^2}{(8 \pi^2-\lambda) \Lambda^2}}\,.
\label{eq_renp}
\eea
 The
first line \eqref{eq_renpa} reports the general finding
of  \cite{Cheung:2024wme}, which we specialise in the second
line \eqref{eq_renp} to our potential of eq \eqref{exphila} (under  our
 hypothesis $\lambda<2$). The function
$W(\phi)$ is given by
\bea
W(\phi)&=&\frac{3}{2} + 128 \pi^4 + 2 \lambda + 2 \lambda^2 - 16 \pi^2 (1 + 2 \lambda)
\nonumber
\\
&
+& \frac{128 \pi^4 (3 - 16 \pi^2 + 2 \lambda) }{(8 \pi^2 - \lambda) }\,
\frac{\phi^2}{\Lambda^2}
+ \frac{8192 \pi^8 }{(8 \pi^2-\lambda )^2 }\,
\frac{\phi^4}{\Lambda^4}\,.
\eea
 We represent the renormalized
 potential in Fig.~\ref{fig_pot2}, right panel.
 Interestingly, the overall shape of the 
 potential \eqref{exphila} is preserved: the flat region
 around the maximum at the origin is unchanged, as well as the initial  potential descend,
 whereas  the DE scalar becomes dynamical. Hence, appropriately regularized
 \cite{Cheung:2024wme} power-law divergences do not 
 spoil the properties of the quintessence potential.~\footnote{We could also say this is called a {\it bounded DE scenario} because power-law loop corrections do not provide uncontrollable corrections to the scalar dynamics.}
 This makes
 a dark energy quintessence scenario 
 based on \eqref{exphila}  
 technically natural from an effective field
 theory viewpoint.
 Only the minimum of the potential gets uplifted, from Minkowski to de Sitter~\footnote{   
 This behavior has  analogies  with  recent results on logarithmic loop corrections and de Sitter vacuua, as  discussed in \cite{Antoniadis:2019rkh}.}. 

Technically, the result \eqref{eq_renp} can be understood as a consequence of an approximate  shift
symmetry of the potential in the flat region
around the maximum, which is broken  by higher
order terms in the potential which cause its rapid growth. Geometrically, such  shift symmetry breaking is delocalized in field space away from
the origin. 
Loop-induced power-law corrections are renormalized by regulators controlled  by  the {\it true} cut-off scale $\Lambda$ in this scenario. 
The procedure modifies the numerical coefficients in the expansion \eqref{eq_tay} by order-one factors,  while preserving the overall structure of the potential. In particular, loop corrections involve higher derivatives, such as $V'''$ and so on, which are suppressed when evaluated on the flat region near the origin. Therefore they  do not affect the flatness of the potential at the local maximum, but only influence its features
away from the origin.

\subsection{The   top-down viewpoint}
\label{sec_td}

From the top-down perspective, quantum gravity constraints on consistent theories of quantum gravity have been put forward in recent years in the form of a set of conjectures, which aim at delimiting the boundary between those EFT's which can be consistently UV completed -- the landscape --  and those which cannot -- the swampland. Among such 
conjectures, one of the more relevant for cosmology,  imposes constraints on the 
derivatives of the scalar potential, as the  de Sitter conjecture \cite{Garg:2018reu, Ooguri:2018wrx}. This conjecture demands  that the scalar potential of the string moduli (describing sizes, shapes and positions in the extra dimensions, and the string coupling) arising from a compactification to four dimensions  should satisfy \cite{Garg:2018reu,Ooguri:2018wrx}:
\be\label{sdSC}
\frac{\sqrt{\nabla^a V\nabla_a V}}{V} \geq \frac{c}{\Mp}  \qquad {\text{or}} \qquad 
  \frac{{\rm min} (\nabla^a \nabla_b V)}{V} \leq -\frac{c'}{\Mp^2}  \,,
\ee
where ``min()" denotes the minimal eigenvalue, and $c$ and $c'$ $ {\mathcal O}(1)$ positive constants.  Whilst there exist physical arguments for these  inequalities to hold in asymptotic regions of the moduli space \cite{Ooguri:2018wrx} -- where large moduli correspond to weak couplings in the corresponding perturbative expansions -- the conjecture speculates that it holds everywhere in moduli space.  
Conjecture \eqref{sdSC} rules out a metastable de Sitter vacuum as the explanation for DE, as we can not have simultaneously $\nabla_a V =0$, $V>0$, and ${\rm eigenvalues\,\,}(\nabla^a \nabla_b V) >0$. On the other hand, recent cosmological data would also seem to indicate that a cosmological constant is not the source the current universe's accelerated expansion. 
The  main alternative, from both theoretical and observational arguments, is {possibly} quintessence (see also \cite{Olguin-Trejo:2018zun}).  Moreover, it is  natural to expect that quintessence candidates are found amongst  string moduli so long as their potential  satisfies \eqref{sdSC}. 

\smallskip

 Potentials of the form \eqref{exphila} belong to the hilltop class and thus for suitable parameter values will satisfy the constraints \eqref{sdSC}. Concretely the quartic and octic potentials \eqref{eq_qpot}, \eqref{eq_oct} have $V''_0/V_0=-4/\Lambda^2$ and thus for suitable $\Lambda$ both satisfy \eqref{sdSC} near the hilltop. The hilltop exponential, while having a more abrupt rise after the minimum, is flatter at the hilltop as seen  from Fig.~\ref{fig_pot2}, with $V''_0/V_0$ given in \eqref{eq:VppVExp}. We shall discuss  in Section \ref{sec_ml} how we can use machine learning  techniques with the theoretical input of the mirage cut-off to obtain examples of scalar potentials that score better in terms of the  curvature at the hilltop. 
 
 We also point out that   the abrupt rise of the potential at finite distances in field space can have interesting relations  with  interpolating domain wall configurations arising in the context of the dynamical cobordism, as   discussed in \cite{Buratti:2021fiv}. Mirage cut-offs may indicate nontrivial constraints on effective field theories,  imposing a fundamental energy scale beyond which new physics must emerge.

Finally, string theory constructions can geometrically  explain  the smallness of the parameter $\epsilon$ in potentials of the form \eqref{eq_tay}. For example, a small overall value for $\epsilon$ can arise from warping in moduli space, when identifying  the scalar field  with an axion  in  axion monodromy inflation \cite{McAllister:2008hb}. We leave additional  work on explicit string-motivated model building to 
future publications.

\subsection{Symmetry breaking:  screening mechanism and initial conditions}
\label{sec_sb}

The bounded dark energy potentials
discussed in Section \ref{sec_bu}
have  a shape resembling symmetry-breaking, Mexican-hat 
configurations -- see Fig.~\ref{fig_typ}.
We now discuss how such feature can be useful
for a DE set-up, since it suggests ways
to develop screening mechanisms, and to
set initial conditions for the field dynamics when
coupling the DE scalar to matter fields.

By promoting the scalar from a real to a complex
 field, we can write $\phi^2\,=\,\Phi^\star \,\Phi$, and the potential becomes a deformed sombrero 
 with a local maximum at the origin and a family of minima at vanishing potential,  for values $\Phi =  \bar \phi e^{i \varphi}$, with 
\( \bar \phi\) the location of the minimum and \( \varphi \)
 an
  arbitrary phase. Concretely, the representative plot
  of Fig.~\ref{fig_typ} becomes a three-dimensional
  figure of revolution, representing a gauge symmetry breaking potential of the typical type considered in field theory setup. 

 In this context, long-range forces associated with the flat
 potential at the origin can be suppressed by means
 of a
 symmetron screening mechanism  \cite{Hinterbichler:2010es}.  
  In essence, the symmetron mechanism   hypothesizes that the field is also coupled to the energy density \(\rho\) of its environment, for example by a coupling proportional to $\rho |\Phi|^2$. When the energy density is large, the field is confined near $|\Phi| \simeq 0$ and acquires a large mass, rendering it unobservable in dense environments such as those on solar system scales, because it is unable to mediate long-range forces. In contrast, in a lower-density environment, the field moves toward its true minimum by rolling down its potential, and the DE dynamics takes place. 
  
    The very same idea can
  be used to  theoretically motivate the initial conditions for the  scalar field
  in our framework. To realise the thawing quintessence
  set-up and make DE dynamical, the scalar
  should start rolling around the tip of the hilltop potential.
   If, in the very early 
  universe, the total energy density was large, a coupling proportional $\rho |\Phi|^2$ make the origin of the scalar
  potential a minimum instead of a maximum, initially keeping the scalar at that location. Only in more recent
  cosmological times, when $\rho$ reduces its value, the effective   scalar potential acquires the 
  sombrero shape of Fig.~\ref{fig_typ}, and the scalar starts rolling away from the origin. 

The aforementioned ideas exploit the possibility that DE is coupled to dark matter and form  an interacting DE-DM system.  Couplings should be chosen appropriately to avoid spoiling the stability of the setup under loop corrections, as discussed in Section \ref{sec_bu}.
The realization of this system and the 
study of its phenomenology  go beyond the scope of this work, and we leave them  to a future
publication.

\section{Comparing  with data}
\label{sec_costest}

We now apply  the theoretical considerations developed in the previous section using the most recent cosmological data. 
In Section \ref{sec_modc2},
data
will determine the best values for the parameters characterising
the models analyzed
in Section \ref{sec_bu}. 
Then, in Section
\ref{sec_ml} we employ machine learning techniques to refine the theoretical framework and leverage real data to identify the most compelling bounded dark energy scenarios.

\medskip
To compare
with observations,
we modify the cosmological Boltzmann code \texttt{CAMB}~\cite{Lewis:1999bs} to implement the models described above. For each scenario, we explore the parameter space through a Markov Chain Monte-Carlo (MCMC) analysis. 
 We vary three model specific parameters alongside the baseline cosmological parameters \mbox{$\{\Omega_{\rm b}h^2,\Omega_{\rm c}h^2,H_0,\tau,A_s,n_s\}$}, adopting wide uniform priors. The model parameter $\Lambda$ is sampled in Planck units and $H_0$ in units of km/s/Mpc. We fix the overall potential amplitude to match the current dark energy scale. We use the following datasets:
 \begin{enumerate}
    \item{CMB from \textit{Planck}:}
\begin{itemize}
    \item[-] \textit{Planck} 2018 low-$\ell$ temperature and polarisation likelihood~\cite{Aghanim:2019ame}.
    \item[-]  \textit{Planck} high-$\ell$ CamSpec TTTEEE temperature and polarization likelihood using \texttt{NPIPE} (\textit{Planck} PR4) data~\cite{Rosenberg:2022sdy}.
    \item[-] \textit{Planck} 2018 lensing likelihood~\cite{Aghanim:2018oex}.
\end{itemize}
 Hereafter, we collectively refer to all the \textit{Planck} CMB likelihoods as `CMB'.
    \item BAO likelihoods from DESI DR1~\cite{DESI:2024lzq,DESI:2024mwx,DESI:2024uvr} 
     \item  Pantheon+~\cite{Brout:2022vxf}, Union3~\cite{Rubin:2023ovl} and DES-Y5~\cite{DES:2024tys} type Ia supernovae likelihoods. 
    
\end{enumerate}

The choice of these datasets is based on the recent findings from the DESI BAO analysis~\cite{DESI:2024mwx}, which suggests a preference for dynamical dark energy with relatively rapid evolution in the recent past, both independently and in combination with supernova datasets from Pantheon+~\cite{Brout:2022vxf}, Union3~\cite{Rubin:2023ovl} and DESY5~\cite{DES:2024tys}.

We employ the MCMC sampler ~\cite{Lewis:2002ah,Lewis:2013hha} in \texttt{Cobaya}~\cite{Torrado:2020dgo}. The MCMC chains are considered to have converged when the Gelman-Rubin diagnostic satisfies the condition $R-1=0.05$. To visualize the constraints and posterior distributions for each model, we use the~\texttt{GetDist} package~\cite{Lewis:2019xzd}. Additionally, we run the Scipy~\cite{2020SciPy-NMeth} minimizer via \texttt{Cobaya} to determine the maximum likelihood point and the corresponding $\chi^2$ values. In Section
\ref{sec_modc2} we present the results for the DE model parameters and the 
$\Lambda$CDM background parameters. The results for the full set of parameters can be found in Appendix~\ref{app:fullmcmc}.  

\subsection{Comparing the models of section \ref{sec_th}
against data}
\label{sec_modc2}
We now find the parameter values for
the models of Section \ref{sec_th} which better fit current data.


\subsubsection*{Hilltop quartic Higgs potential}

\begin{figure}[H]
    \centering
    \includegraphics[width=0.75\linewidth]{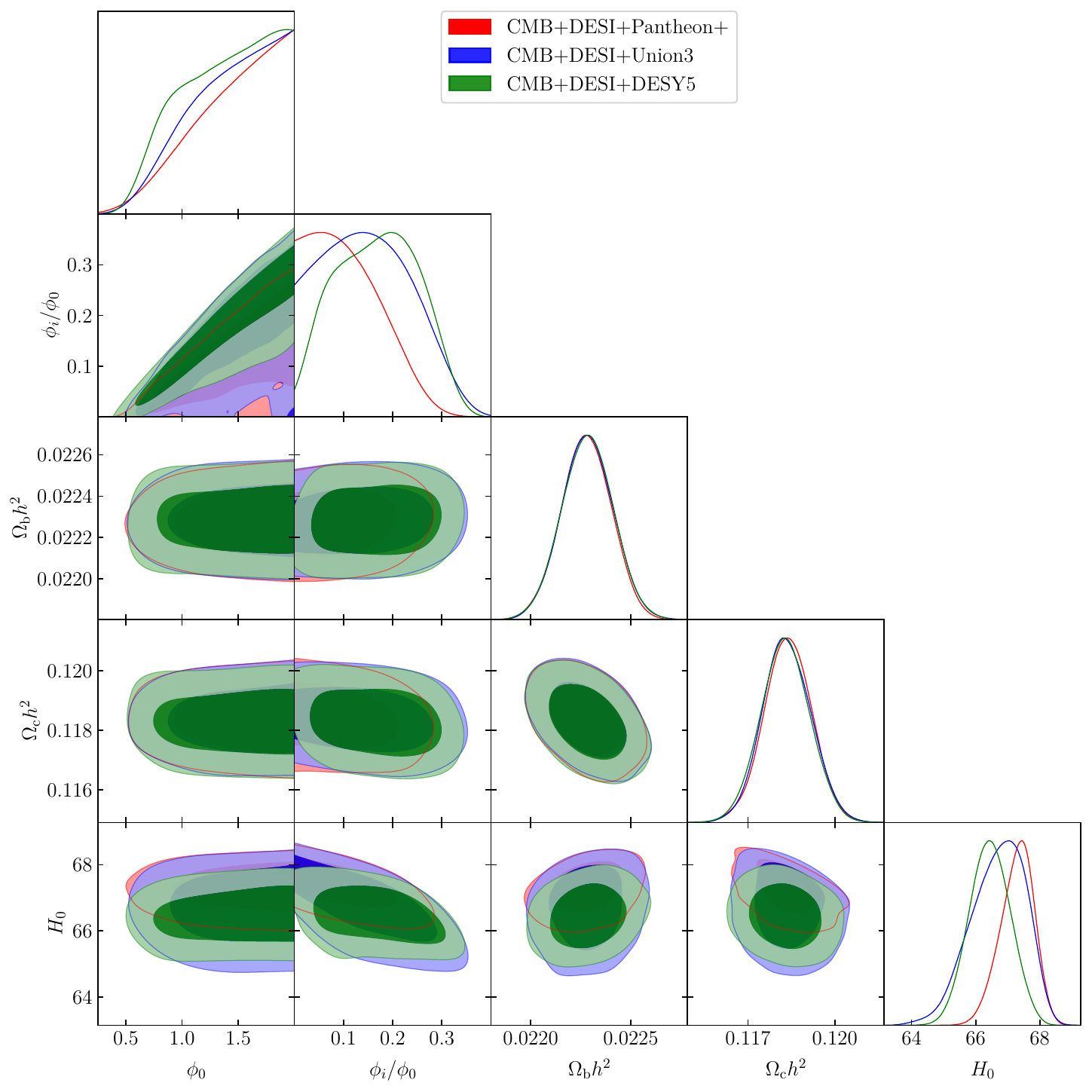}
    \caption{Constraints on the quartic Higgs hilltop model of
    eq \eqref{eq_qpot} ($68\%$ and $95\%$ contours). }
    \label{fig:Higgs_mcmc0}
\end{figure}

We start presenting  the results for the hilltop quartic model analysed in~\cite{Bhattacharya:2024kxp}, to better make the comparison between our hilltop models. The sampled parameters are $\Lambda$
 expressed
in Planck units~\footnote{From now on we use
Planck units  to express the cut-off value
$\Lambda$.} and the initial relative field value $\phi_i/\Lambda$, which represent the steepness of the potential and the initial displacement from the maximum, respectively. The results are shown in Figure \ref{fig:Higgs_mcmc0}, while Table \ref{tab:param_limits_Higgs} summarises the parameter means and the $68\%$ limits.  The resulting value of $\Lambda$
is of the order of the Planck scale. 
As discussed previously in~\cite{Bhattacharya:2024kxp}, we notice that for a steeper potential, i.e.~a smaller $\Lambda$, the allowed values of $\phi_i/\Lambda$ are squeezed into a smaller region around the origin. Conversely, a shallower potential allows for a broader range but away from $\phi_i/\Lambda\approx 0$. This constraint arises to ensure the presence of dynamical dark energy in the present epoch, in particular for the CMB+DESI+DESY5 dataset shown in green.
Note that in this case we varied two model parameters, whereas the octic and exponential hilltop models include an additional parameter (respectively $c_0$ and $\lambda$) that should also be sampled. 
The mean value for the cut-off
$\Lambda$ is of the order 
of the Planck scale, while
the dark energy scale $V_0 = \epsilon \Lambda^4$ is around
120 orders of magnitude smaller. This in fact demonstrates
that in our seup a small value
for $\epsilon$ is very well motivated. 

\renewcommand{\arraystretch}{1.45}
\begin{table}[H]
    \centering
\begin{tabular} { |c| c| c| c|}

\hline
\rowcolor{gray!30} 
 \bf{Parameter} &  {\bf +Pantheon+} & {\bf +Union3} &  {\bf +DESY5} \\
\hline
{\boldmath$\Lambda         $} & $> 1.29                    $ & $> 1.24                    $ & $> 1.17                    $\\

{\boldmath$\phi_i/\Lambda  $} & $< 0.142                   $ & $0.151^{+0.073}_{-0.12}    $ & $0.169\pm 0.081            $\\

{\boldmath$\Omega_\mathrm{c} h^2$} & $0.11838\pm 0.00082        $ & $0.11835\pm 0.00084        $ & $0.11829\pm 0.00084        $\\

{\boldmath$\Omega_\mathrm{b} h^2$} & $0.02228\pm 0.00012        $ & $0.02228\pm 0.00013        $ & $0.02228\pm 0.00013        $\\

{\boldmath$H_0            $} & $67.29^{+0.59}_{-0.45}     $ & $66.7^{+1.0}_{-0.70}       $ & $66.44\pm 0.64             $\\

$\phi_i                    $ & $0.174^{+0.071}_{-0.17}    $ & $0.235^{+0.088}_{-0.23}    $ & $0.26^{+0.10}_{-0.24}      $\\
\hline
\end{tabular}
    \caption{Quartic hilltop model: parameter means and $68\%$ limits for the addition of the different supernovae datasets to the CMB+DESI combination. }
    \label{tab:param_limits_Higgs}
\end{table}

\subsubsection*{Hilltop octic potential}

\begin{figure}[H]
    \centering
    \includegraphics[width=0.75\linewidth]
    {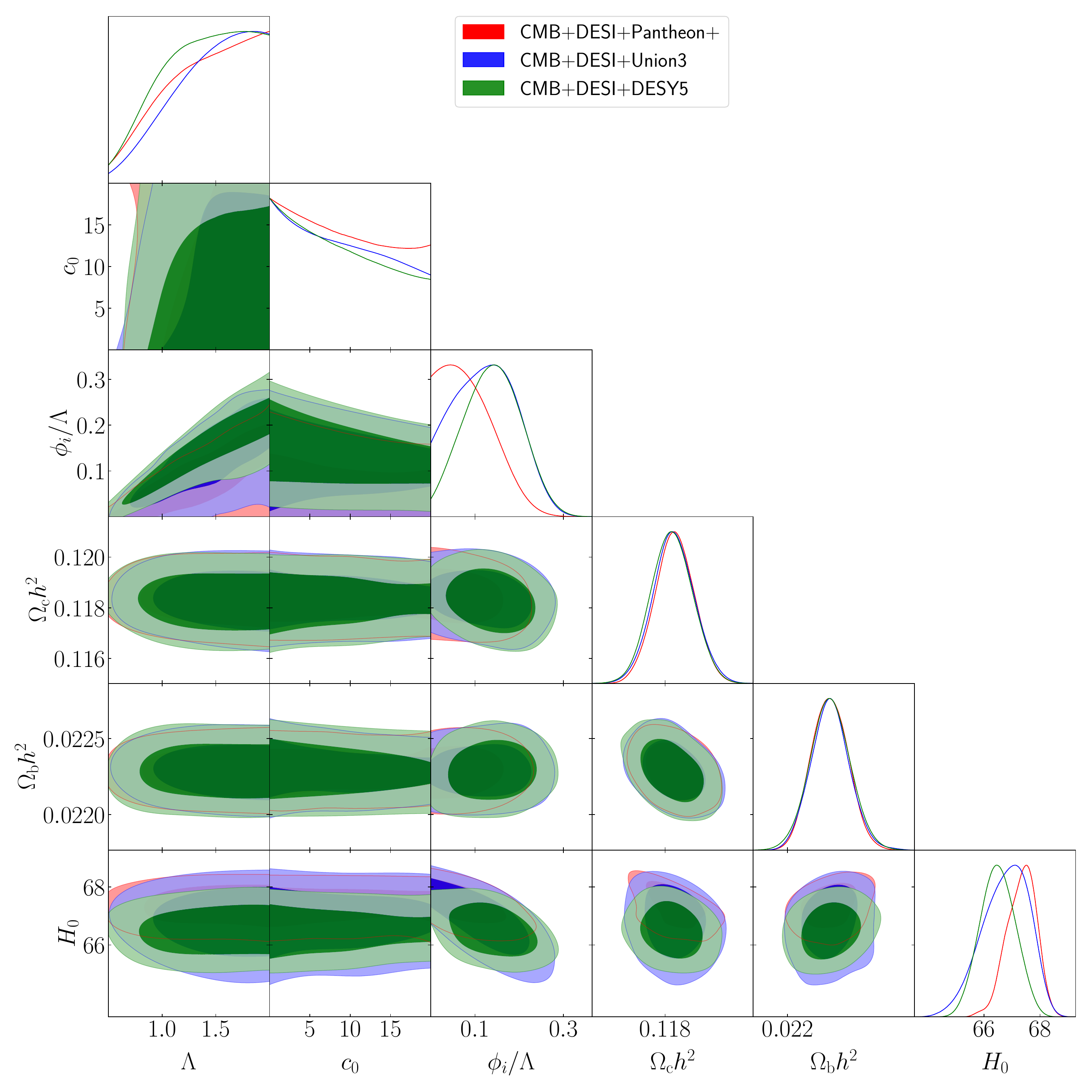}
    \caption{Parameter constraints ($68\%$ and $95\%$ contours) for the hilltop octic model of eq \eqref{eq_oct}.}
   \label{Octic_triangle}
\end{figure}

\begin{figure}[H]
    \centering
        \includegraphics[width=0.5\linewidth]{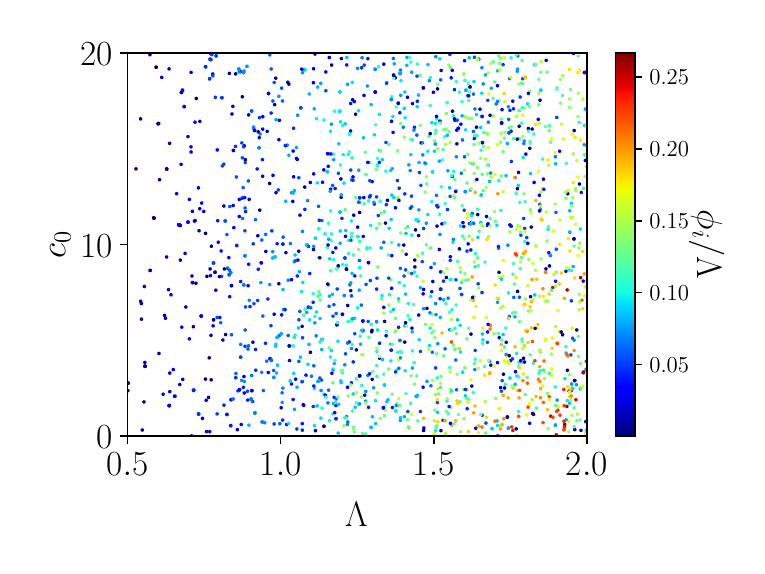}\hspace{-0.5cm}
        \includegraphics[width=0.5\linewidth]{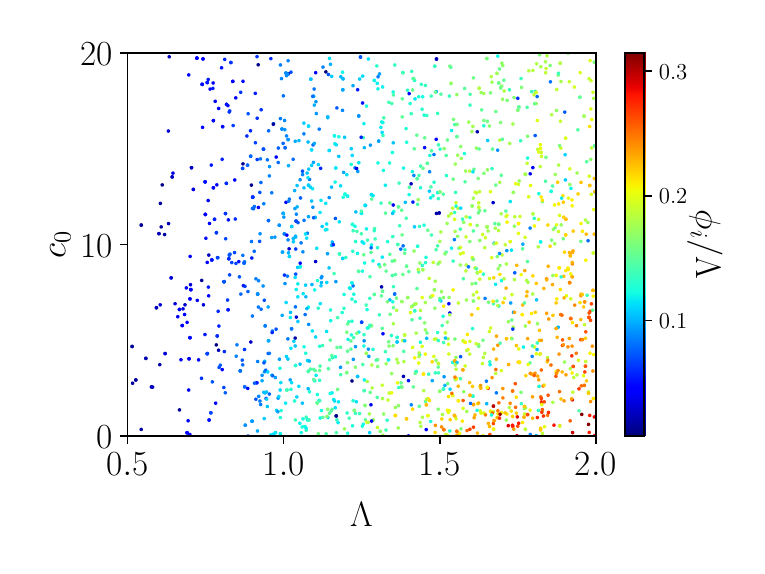}
    \caption{The DE model parameters $c_0,\phi_i/\Lambda,\Lambda$ plotted as a 3D scatter-plot generated from the MCMC samples for the octic model, using the CMB+DESI+Pantheon+ dataset (left) and CMB+DESI+DESY5 (right). }
    \label{fig:3D_Octic}
\end{figure}

For the octic model, we sample the parameters $\Lambda$, $c_0$ and $\phi_i/\Lambda$. As before, $\phi_i/\Lambda$ represents the initial field displacement from the maximum, in units of $\Lambda$, while $\Lambda$ and $c_0$ control the steepness of the potential. More specifically, larger values of $c_0$ and smaller values of $\Lambda$ result in a steeper potential, and vice versa. The resulting parameter posterior distributions are plotted in Figure \ref{Octic_triangle}, with the $68\%$ limits presented in Table \ref{tab:param_limits_oct}. We learn that present data is unable to provide meaningful constraints on the parameter $c_0$, implying that the corresponding term in the scalar field potential does not play a significant role in the field evolution. Note however, that for larger $c_0$, a large initial field displacement $\phi_i/\Lambda$ is excluded (otherwise the potential becomes too steep). Given these considerations, it is not surprising that the $\phi_i/\Lambda$ vs $\Lambda$ constraints resemble those of the quartic model. These features are also reflected in the three-dimensional scatter plot of the three parameters $c_0,\phi_i/\Lambda,\Lambda$, shown in Figure~\ref{fig:3D_Octic} for the CMB+DESI data in combination with Pantheon+ and DESY5.

\smallskip 
\noindent



\renewcommand{\arraystretch}{1.4}
\begin{table}[H]
    \centering
\begin{tabular} { |c| c| c| c|}

\hline
\rowcolor{gray!30} 
 \bf{Parameter} &  {\bf +Pantheon+} & {\bf +Union3} &  {\bf +DESY5} \\
\hline
{\boldmath$\Lambda              $} & $> 1.21                   $ & $> 1.29                   $ & $> 1.18      $\\

{\boldmath$c_0              $} &  ---                    & $< 12.4                   $ & $< 11.8    $\\

{\boldmath$\phi_i/\Lambda       $} & $< 0.113$                          & $0.127\pm 0.067  $                        & $0.141\pm 0.062                   $\\

{\boldmath$\Omega_\mathrm{c} h^2$} & $0.11839\pm 0.00079        $ & $0.11835\pm 0.00084        $ & $ 0.11828\pm 0.00084     $\\

{\boldmath$\Omega_\mathrm{b} h^2$} & $0.02229\pm 0.00012        $ & $0.02229\pm 0.00013        $ & $0.02229\pm 0.00014       $\\

{\boldmath$H_0            $} & $67.34^{+0.59}_{-0.47}     $ & $66.78^{+0.97}_{-0.70}    $ & $66.50\pm 0.62    $\\

$\phi_i                    $ & $0.136^{+0.044}_{-0.14}         $ & $0.199^{+0.091}_{-0.17}$ & $0.214^{+0.097}_{-0.17}      $\\
\hline
\end{tabular}
    \caption{Hilltop octic model: parameter means and $68\%$ limits for the addition of the different supernovae datasets to the CMB+DESI combination.}
    \label{tab:param_limits_oct}
\end{table}

\subsubsection*{Hilltop exponential potential}

\begin{figure}
    \centering
    \includegraphics[width=0.75\linewidth]
    {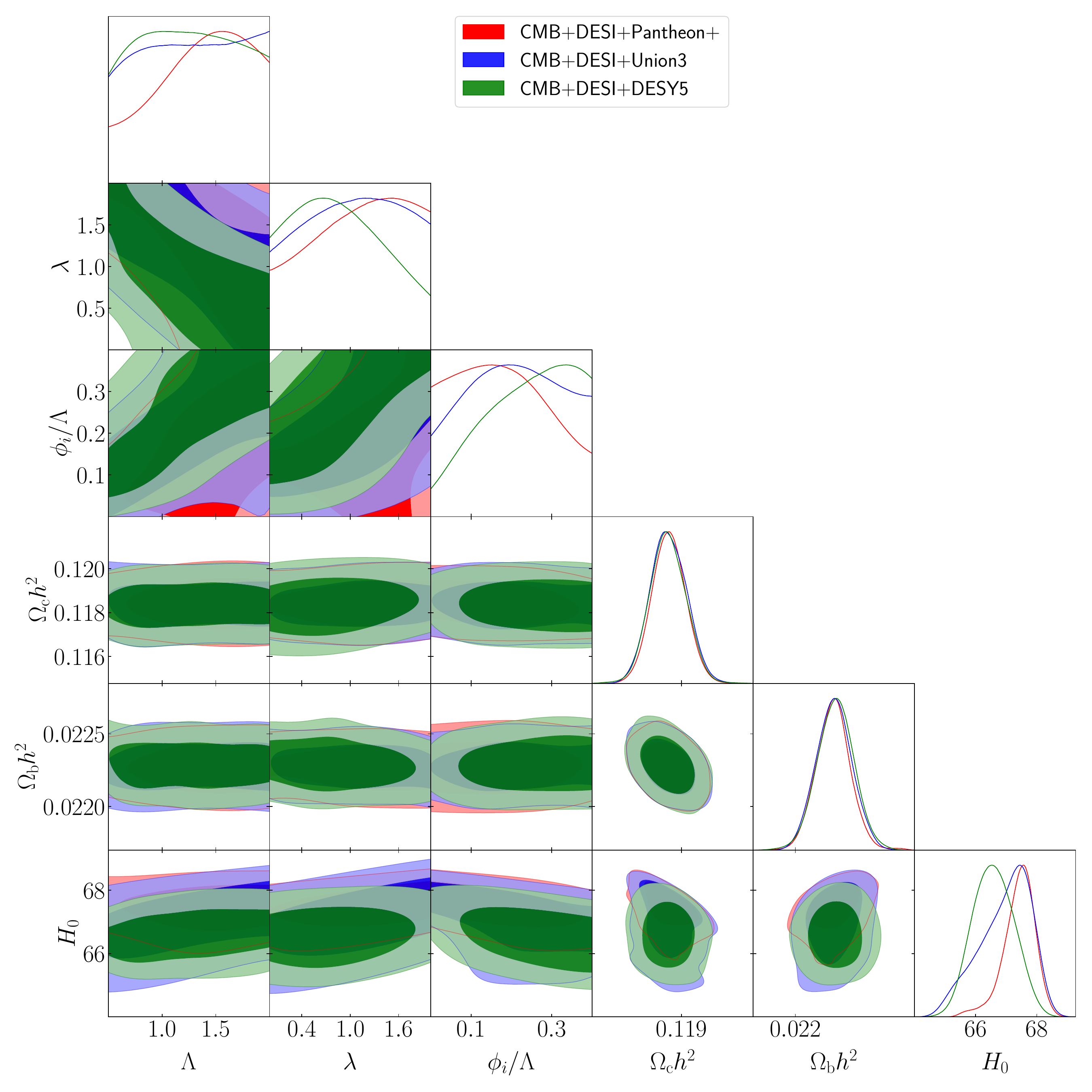}
    \caption{Parameter constraints ($68\%$ and $95\%$ contours) for the hilltop exponential model of eq \eqref{exphila}.}
   \label{Exp_triangle}
\end{figure}

\begin{figure}[H]
    \centering
    \includegraphics[width=0.5\linewidth]{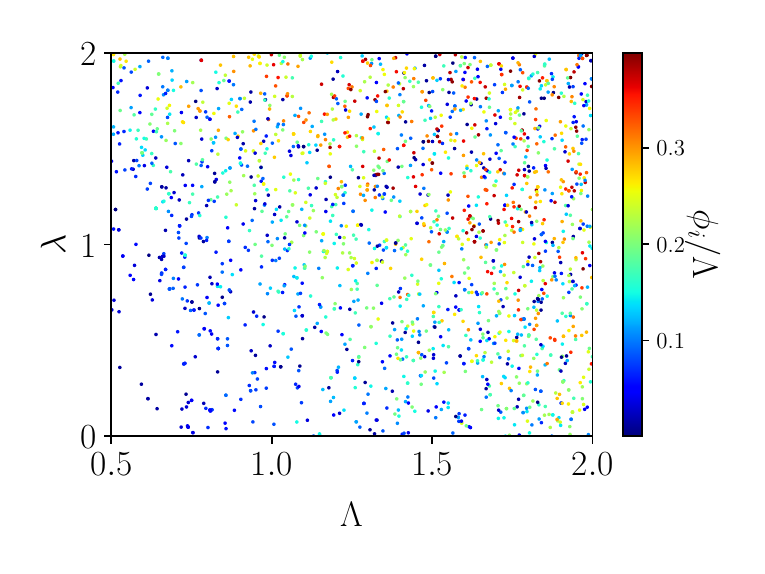}\hspace{-0.5cm}
        \includegraphics[width=0.5\linewidth]{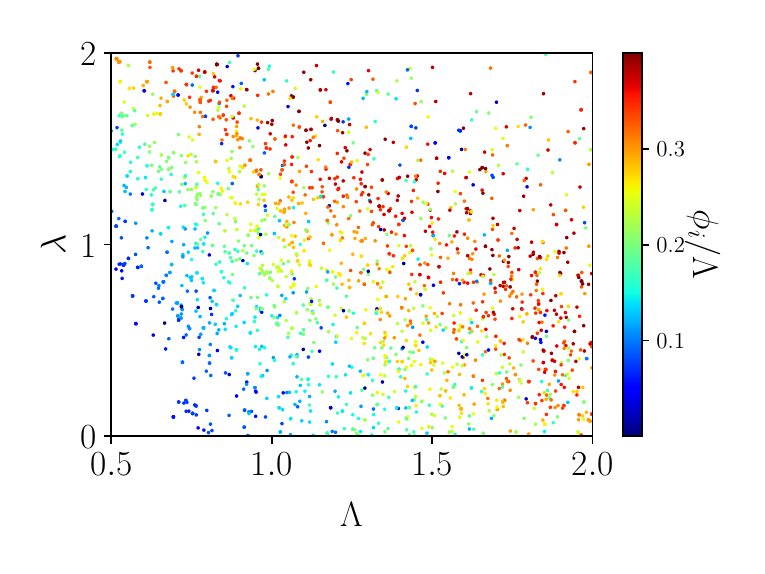}
    \caption{The DE model parameters $\lambda,\phi_i/\Lambda,\Lambda$ plotted as a 3D scatter-plot generated from the MCMC samples for the Hilltop Exponential model, using the CMB+DESI+Pantheon+ dataset (left) and CMB+DESI+DESY5 (right). }    \label{fig:3D_Exp}
\end{figure}

For the hilltop exponential model, we vary the parameters $\Lambda$, $\lambda$ and $\phi_i/\Lambda$. The results are plotted in Figure \ref{Exp_triangle} and the $68\%$ limits presented in Table \ref{tab:param_limits_exp}. Once again, we identify $\phi_i/\Lambda$ as the initial field displacement and we track  the role of the other model parameters, $\lambda$ and $\Lambda$, in controlling the steepness of the potential. Additionally, in this case, the parameter $\lambda$ also influences the flatness of the potential around the origin, with bigger values resulting in a flatter area.  This is reflected in the $\lambda$ vs $\phi_i/\Lambda$ contours, where we learn that for larger $\lambda$ (potential flatter at the origin), the initial field displacement can not be too small, in particular for the DESY5 dataset which requires the largest deviation from $\Lambda$CDM. A similar effect can be noticed in the $\lambda$ vs $\Lambda$ plot where one cannot have a potential flat at the hilltop as well as away from it for the same dataset, otherwise the late-time dark energy behavior resembles that of the cosmological constant. The $\phi_i/\Lambda$ vs $\Lambda$ contours are interpreted similarly to the previous two potentials considered in this section. The 3-dimensional scatter plot of the three parameters $\lambda,\phi_i/\Lambda,\Lambda$, in Figure~\ref{fig:3D_Exp}, reveal the same trends. 

\renewcommand{\arraystretch}{1.4}
\begin{table}[H]
    \centering
\begin{tabular} { |c| c| c| c|}

\hline
\rowcolor{gray!30} 
 \bf{Parameter} &  {\bf +Pantheon+} & {\bf +Union3} &  {\bf +DESY5} \\
\hline
{\boldmath$\Lambda              $} & $1.36^{+0.64}_{-0.20}                   $ & ---                  & --- \\

{\boldmath$\lambda              $} & $> 0.823$                    & --- & $0.89^{+0.36}_{-0.80}   $\\

{\boldmath$\phi_i/\Lambda       $} & $0.180^{+0.068}_{-0.17}$                          & $0.21\pm 0.11  $                        & $> 0.183
                  $\\

{\boldmath$\Omega_\mathrm{c} h^2$} & $0.11840\pm 0.00079        $ & $0.11839\pm 0.00085 $ & $ 0.11835\pm 0.00085   $\\

{\boldmath$\Omega_\mathrm{b} h^2$} & $0.02227\pm 0.00013       $ & $0.02228\pm 0.00013  $ & $0.02229\pm 0.00013    $\\

{\boldmath$H_0            $} & $67.42^{+0.57}_{-0.39}    $ & $67.0^{+1.1}_{-0.60}   $ & $66.58\pm 0.67  $\\

$\phi_i                    $ & $ 0.25^{+0.10}_{-0.23}        $ & $0.28^{+0.12}_{-0.24}$ & $0.31^{+0.14}_{-0.26}     $\\
\hline
\end{tabular}
    \caption{Hilltop exponential model: parameter means and $68\%$ limits for the addition of the different supernovae datasets to the CMB+DESI combination.}
    \label{tab:param_limits_exp}
\end{table}

\subsubsection*{Summary of the results so far}

\begin{figure}[H]
     \centering
     \includegraphics[width=0.49\linewidth]{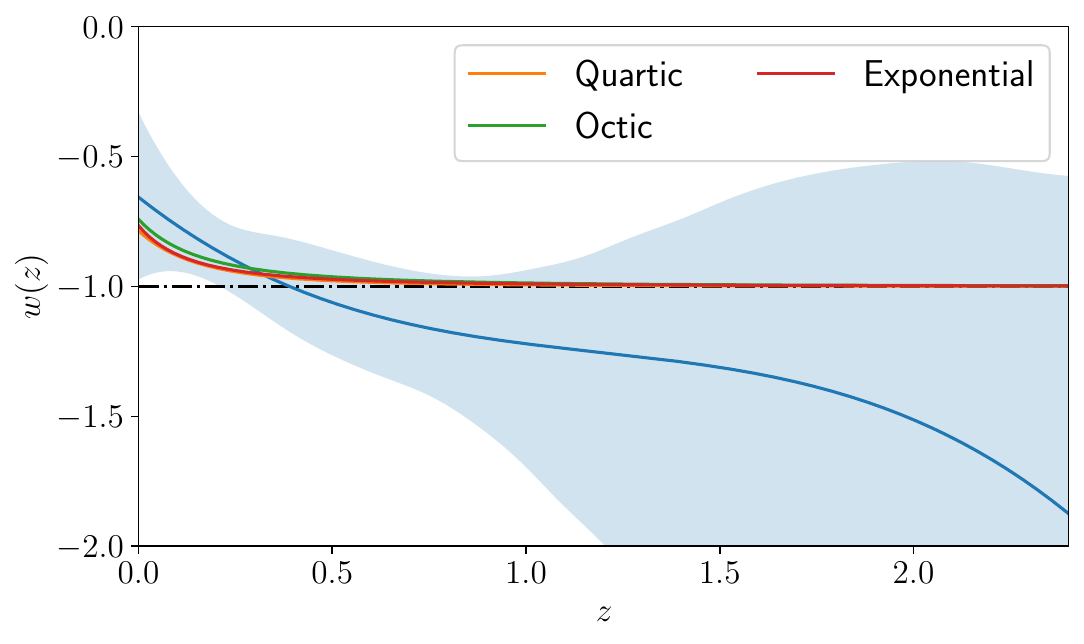}
         \includegraphics[width=0.49\linewidth]{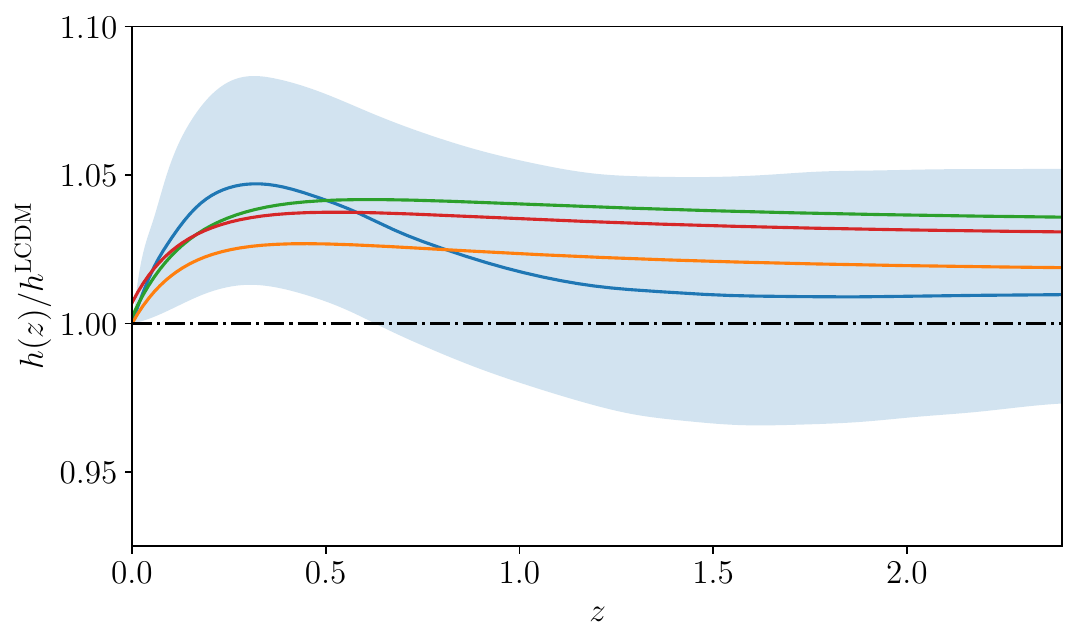}
     \caption{The evolution of  $w_{\rm \phi}(z)$ and $h(z)\equiv H(z)/H_0$ is shown for the best-fit hilltop models
     of section \ref{sec_modc2}, and compared to the DESI reconstruction (blue line) using CMB+DESI+Union3 data~\cite{Calderon:2024uwn}. The shaded regions represent the 95\% confidence regions. As the CPL parametrization closely follows the DESI reconstruction, it is not plotted separately.  At $z=0$, $w_{0} = -0.65$ for the reconstruction, while the values for the hilltop models are: $w_0=-0.74$ for the octic model, $w_0=-0.77$ for the exponential model, and $w_0=-0.79$, for the quartic model.
     The best-fit parameter values for each model can be found in Appendix \ref{app:fullmcmc}.}
     \label{fig:bestfit_v_reconstruction}
\end{figure}

In Fig.~\ref{fig:bestfit_v_reconstruction}
we  visualize some consequences of the
results obtained
so far, by
  plotting the evolution
of the scalar field equation
of state and the Hubble
parameter as a function
of redshift for the three
models
in this section. We compare these quantities
with the  DESI reconstruction using CMB+DESI+Union3 data~\cite{Calderon:2024uwn}, also representing
the corresponding error bars. Our conclusions are as follows:
\begin{itemize}
\item 
The left panel in Fig.~\ref{fig:bestfit_v_reconstruction}
indicates that the models
are in the right track for reproducing the evolution
of the DE equation of state as function of redshift,
although they do not  succeed in matching the DESI reconstruction
exactly. The problem being 
that the shape of the potential
must be tuned very precisely to allow for a scalar dynamics that better fits
observations. 
\item
Moreover, Tables \ref{tab:param_limits_Higgs}-\ref{tab:param_limits_exp} show that in the models   considered so far
 the mean values of the  cut-off scales $\Lambda$  
of the order of Planck scale. It would instead be desirable
to find setup with a smaller cut-off scale, in order to ensure they are not destabilized  by corrections involving graviton loops (more on this in Section \ref{sec_out}).
\end{itemize}
 To improve on these points, we now proceed 
using machine learning techniques to build bounded
dark energy potentials starting
from data themselves. Machine learning will be able to find bounded DE models with smaller cut-off scales, and which allow to better fit current observations. 

\subsection{A machine learning approach}
\label{sec_ml}

The results of the previous
Section \ref{sec_modc2} show 
that bounded DE scenarios can 
fit current observational data well,
although they do not reconstruct
DESI results as well as the CPL
parametrization at low redshifts:
see Fig.~\ref{fig:bestfit_v_reconstruction}.
It is interesting to apply machine learning (ML) techniques
to reconstruct good examples
of bounded DE potentials that
better fit data, possibly with smaller values of the model
cut-off.

\begin{figure}[H]
    \centering
    \includegraphics[width=0.49\linewidth]{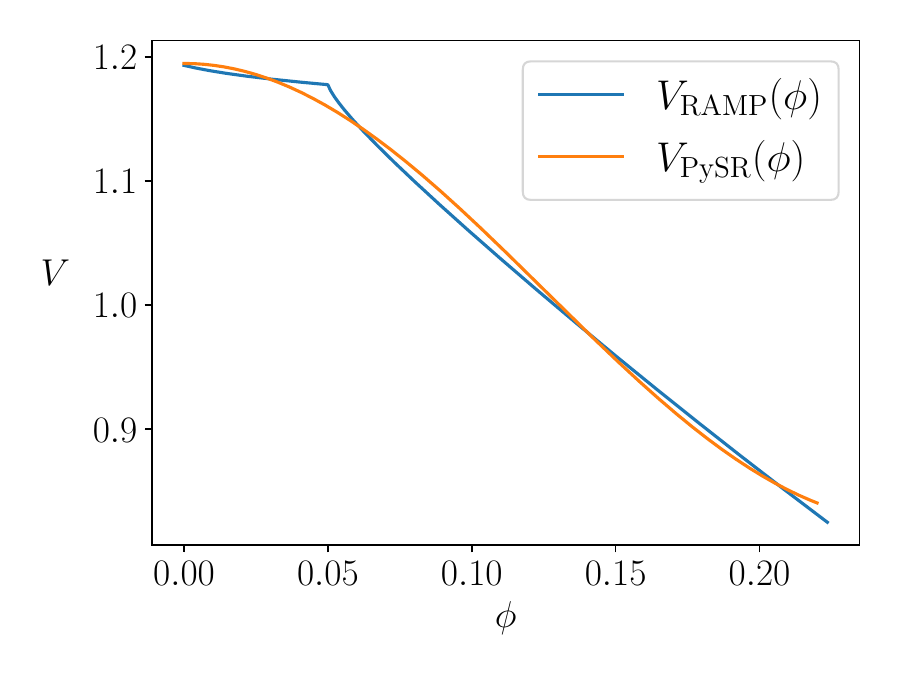} \includegraphics[width=0.49\linewidth]{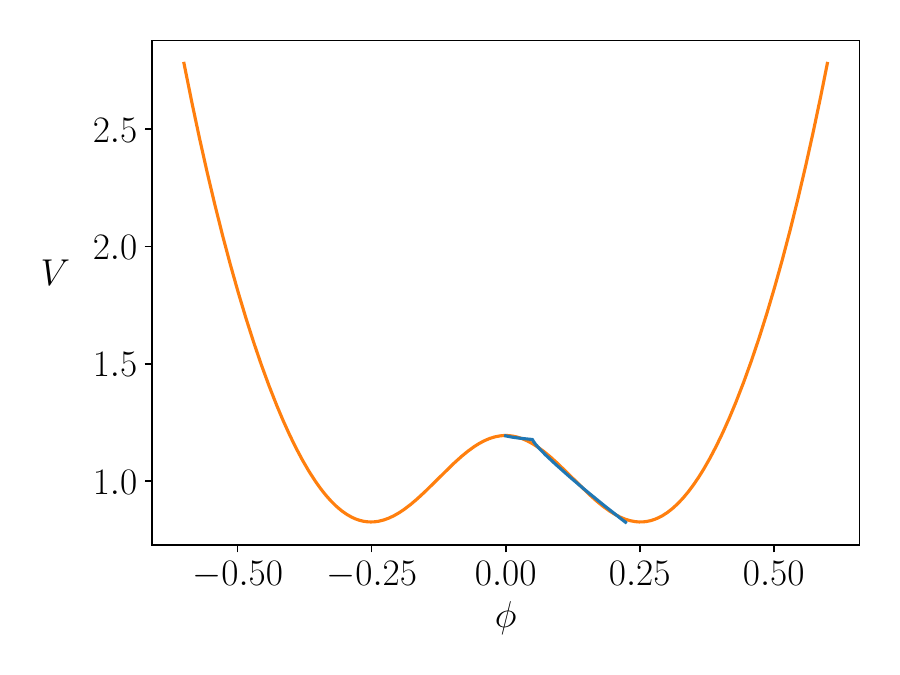}
    \caption{
    {\bf Left panel:} PySR fit (orange line) of the ramp model of \cite{Notari:2024rti} (blue line). {\bf Right panel:} Representation of the PySR
    potential \eqref{V_pysr} for a more extended interval of $\phi$. }
    \label{fit}
\end{figure}

We make use of techniques
based on 
 symbolic regression (SR), capable of identifying interpretable symbolic expressions that are translated
into scalar potentials.
The method balances prediction accuracy and model complexity.  We aim to find a hilltop-type potential, in the
same
class of   models studied in 
the theory Section \ref{sec_th},
which mimic a cosmological constant at early times, while better capturing the steep rise
of 
in the DESI reconstruction of $w(z)$ at low redshifts -- see  Fig.~\ref{fig:bestfit_v_reconstruction}, blue line. For this purpose, 
as benchmark starting point,  we consider a CPL-like model that never crosses the phantom divide at $w = -1$. We follow the example of \cite{Notari:2024rti}, who reverse engineer $w(z)$ to find a potential. We start assuming $w={\rm max}\left[-1, w_0+w_a(1-a)\right]$, where $w_0=-0.65$ and $w_a=-1.23$, which lead
to the so-called ramp model in \cite{Notari:2024rti}. (The ramp
potential $V_{\rm RAMP}(\phi)$ is represented in Fig.~\ref{fit}, left panel.) 
Taking inspiration from \cite{Sousa:2023unz,Sousa-Neto:2025gpj}, we then
 apply SR to regularize the potential features, and to  determine a smooth hilltop potential symmetric around the origin,  
 rapidly increasing  at large values of $\phi$. 

We use the Python package PySR~\cite{cranmer2023interpretablemachinelearningscience} to explore the model space under a set of constraints
on the allowed  functional form
for the desired scalar potential. We use the following set of operators  $\{+,-,*,/,{\rm power},\exp\}$, and for simplicity 
we further restrict the search space by avoiding nested composition of operators, such as $e^{e^{e\phi}}$. 
In particular, we are interested in a generalization of the hilltop exponential potential \eqref{exphila}. Therefore, we specify a custom template, $V_{\rm PySR}(\phi)=V_0 \,e^{\lambda\,\phi^2/\Lambda^2}f(\phi^2)$, allowing PySR to determine the optimal form $f(\phi^2)$ and $\lambda$, i.e. the formula that optimally balances accuracy and simplicity~\footnote{The
corresponding codes are available upon request.
}. 

Since PySR is an evolutionary algorithm, different runs can yield different expressions. Such behaviour is expected when considering the numerous possible combinations of the allowed operators, even under the constraints we imposed. We do however, note recurring patterns in the generated equations, which consistently approximate the ramp potential. We report a particularly well-performing result, which resembles the form of the hilltop models studied in this paper, and we test it against data, as in Section \ref{sec_modc2}. The resulting potential is 
\begin{equation}
    V_{\rm PySR}(\phi)\,=\,V_0\,e^{0.01\,\phi^2/\Lambda^2}\left(a\,\frac{\phi^2}{\Lambda^2}+b+e^{c\phi^2/\Lambda^2}\right),
    \label{V_pysr}
\end{equation}
with best-fit coefficients $a=0.04845719$, $b=0.19458601$ and $c=-0.20447332$. We set $\Lambda=0.1$ (well below the Planck scale), and $V_0$ at the value
of the DE scale. As earlier, we express dimensionful quantities in Planck units.  The potential is shown in Figure \ref{fit}. In the left panel, we zoom the
potential around the origin showing that it
represents a smooth version of
the ramp model of \cite{Notari:2024rti},
without fully capturing its sharp feature occurring at $\phi\approx0.06$. The reason being that a tighter fit of
the ramp scenario would require a higher complexity of the expression, which is not rewarded by a much higher accuracy. Said this, as shown in Figure \ref{bestfitw} the ML approach succeeds in reproducing a better trend of $w(z)$ at late times compared to the other hilltop models inspected in the previous sections. {Figure \ref{bestfitw} shows the evolution of $w(z)$ and $h(z)$, obtained using the results of a minimization process, conducted as in Section \ref{sec_modc2}. Using the same datasets -- CMB+DESI+Union3~\cite{Calderon:2024uwn} -- we varied the usual baseline cosmological parameters \mbox{$\{\Omega_{\rm b}h^2,\Omega_{\rm c}h^2,H_0,\tau,A_s,n_s\}$} along with a single additional parameter, $\phi_i$, the initial
position
for the scalar field -- see Appendix \ref{app:fullmcmc} for full results.  Since we left SR
to reconstruct the potential parameters
from the data, we do not vary them. 
Hence, unlike the previous models, in this case the steepness of the potential is set by the best-fit values found by PySR.

\begin{figure}[H]
    \centering
    \includegraphics[width=0.45\linewidth]{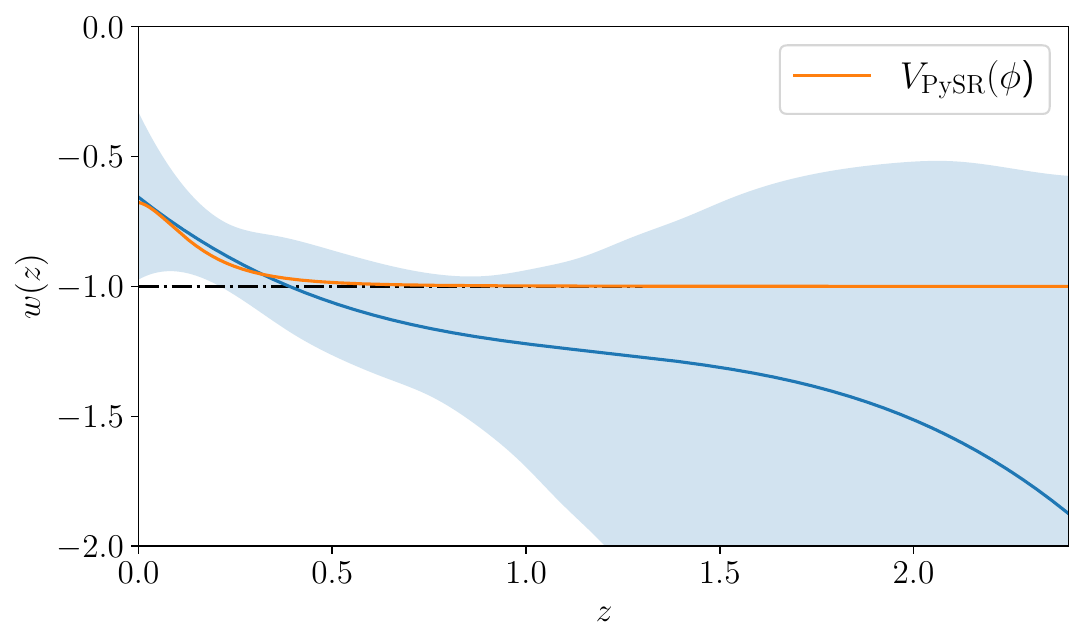} \includegraphics[width=0.45\linewidth]{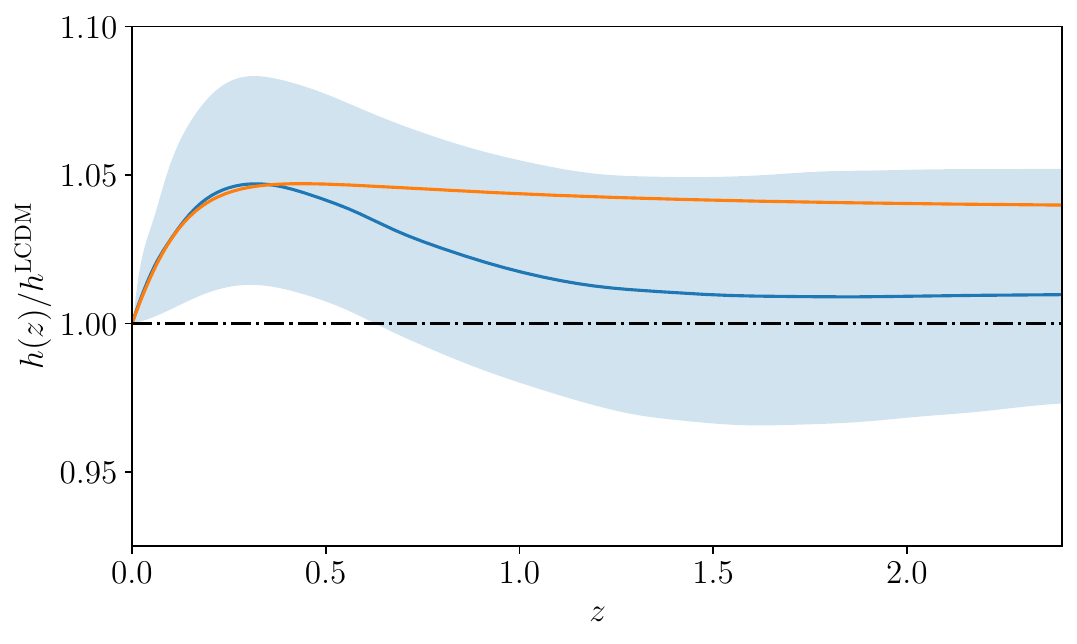}
    \caption{The figure shows the evolution of  $w_{\rm \phi}(z)$ and $h(z)\equiv H(z)/H_0$ for the potential determined  by a ML method
    based on PySR, compared to the DESI reconstruction (blue line) using CMB+DESI+Union3 data~\cite{Calderon:2024uwn}. At $z=0$ we set $w_{0} = -0.65$ and $w_{0} = -0.68$ for the reconstruction and the best-fit hilltop model, respectively. {The full set of best-fit parameters for $V_{\rm PySR}$ can be found in Appendix \ref{app:fullmcmc}.}}
    \label{bestfitw}
\end{figure}

{To  test how well the potential generated by PySR fits the data, we compute the $\Delta\chi^2$  comparing this model with $\Lambda \rm{CDM}$,  CPL, and the quartic, octic and exponential  models of Section \ref{sec_th}. The results for the CMB+DESI+Union3 combination are:
\begin{align}
    \Delta\chi^2_{\rm PySR}= \chi^2_{\rm PySR}-\chi^2_{\Lambda\rm{CDM}}= -4.6 \\ \Delta\chi^2_{\rm PySR}= \chi^2_{\rm PySR}-\chi^2_{\rm CPL}= 8.5 \\
    \Delta\chi^2_{\rm  PySR}= \chi^2_{\rm PySR}-\chi^2_{\rm quartic}= -2.2 \\
    \Delta\chi^2_{\rm  PySR}= \chi^2_{\rm PySR}-\chi^2_{\rm octic}= -2.1 \\
    \Delta\chi^2_{\rm  PySR}= \chi^2_{\rm PySR}-\chi^2_{\rm exp}= -1.4
\end{align}
where $\chi^2=-2\rm ln \mathcal{L}_{max}$ for a given model. This shows that the ML model  provides a better fit to the data compared to $\Lambda \mathrm{CDM}$ and the other hilltop potentials, 
 although the phenomenological CPL parametrization remains the most favored scenario by data.

At the theory level, the ML potential \eqref{V_pysr} is interesting since it keeps the cut-off
scale $\Lambda$ well smaller than the Planck scale. Additionally,  it satisfies better the swampland de Sitter conjecture inequalities \eqref{sdSC} with respect
to the other, human-generated models, suggesting  once more that such inequalities are motivated  by
observational results. 

\section{Outlook}
\label{sec_out}

Recent data indicate that dark energy is dynamical. If so, it is crucial to develop  well-controlled theoretical models to understand its nature.
Motivated by these facts,  we presented a model of hilltop quintessence dubbed
bounded dark energy, characterized by a potential which exhibits a hilltop that appears relatively flat compared to its steep  growth for large values of $\phi$. 
We demonstrated that such potentials align closely with the recent proposal of theories characterized by a mirage cut-off \cite{Cheung:2024wme}. We provided arguments supporting the stability of the quintessence potential against large quantum corrections. Additionally, we explored how bounded dark energy models naturally fit within top-down quantum gravity frameworks. Furthermore, we presented general considerations indicating that our hilltop scenarios can exhibit screening mechanisms and proposed methods for selecting appropriate initial conditions for the scalar field.
%
 
 We presented
simple, explicit examples of bounded
dark energy models, confronting them with the most recent cosmological data sets. Although
 the model predictions are in reasonably good agreement with 
 observations,
we identified aspects that could be improved. Accordingly, we employed
a machine learning approach for
designing bounded dark energy potentials that are more consistent with observational results. We  have shown that the resulting models fit data very well. 

It will be interesting to further refine our understanding of bounded dark energy scenarios. While we focused on loop corrections associated with DE scalar field, it would be interesting to also consider the effects of graviton loops. If the model cut-off $\Lambda$ is much smaller than the Planck scale, hopefully such graviton loops provide only negligible corrections. It would also be interesting to further refine the machine learning methods 
to build new potentials with parameters accommodating future observational results, and additionally imposing more stringent and well motivated theoretical priors for the shape
of
 quintessential potentials. 

\subsection*{Acknowledgments}

We thank Sukannya Bhattacharya, Carlos N\'u\~nez and Susha Parameswaran for discussions on related topics. We are partially funded by the STFC grants ST/T000813/1 and ST/X000648/1. We also acknowledge the support of the Supercomputing Wales project, which is part-funded by the European Regional Development Fund (ERDF) via Welsh Government. 
For the purpose of open access, the authors have applied a Creative Commons Attribution licence to any Author Accepted Manuscript version arising. Research Data Access Statement: No new data were generated for this manuscript.

\newpage

\begin{appendix}
\section{Full parameter space constraints}\label{app:fullmcmc}
In this section, we provide the plots and tables containing the complete set of parameters for the three hilltop models {and for the symbolic regression model}, as  discussed in the main text. The results for the Hilltop quartic Higgs potential are obtained from~\cite{Bhattacharya:2024kxp}.

\subsection*{Hilltop quartic Higgs potential}

\begin{figure}[H]
    \centering
    \includegraphics[width=0.95\linewidth]{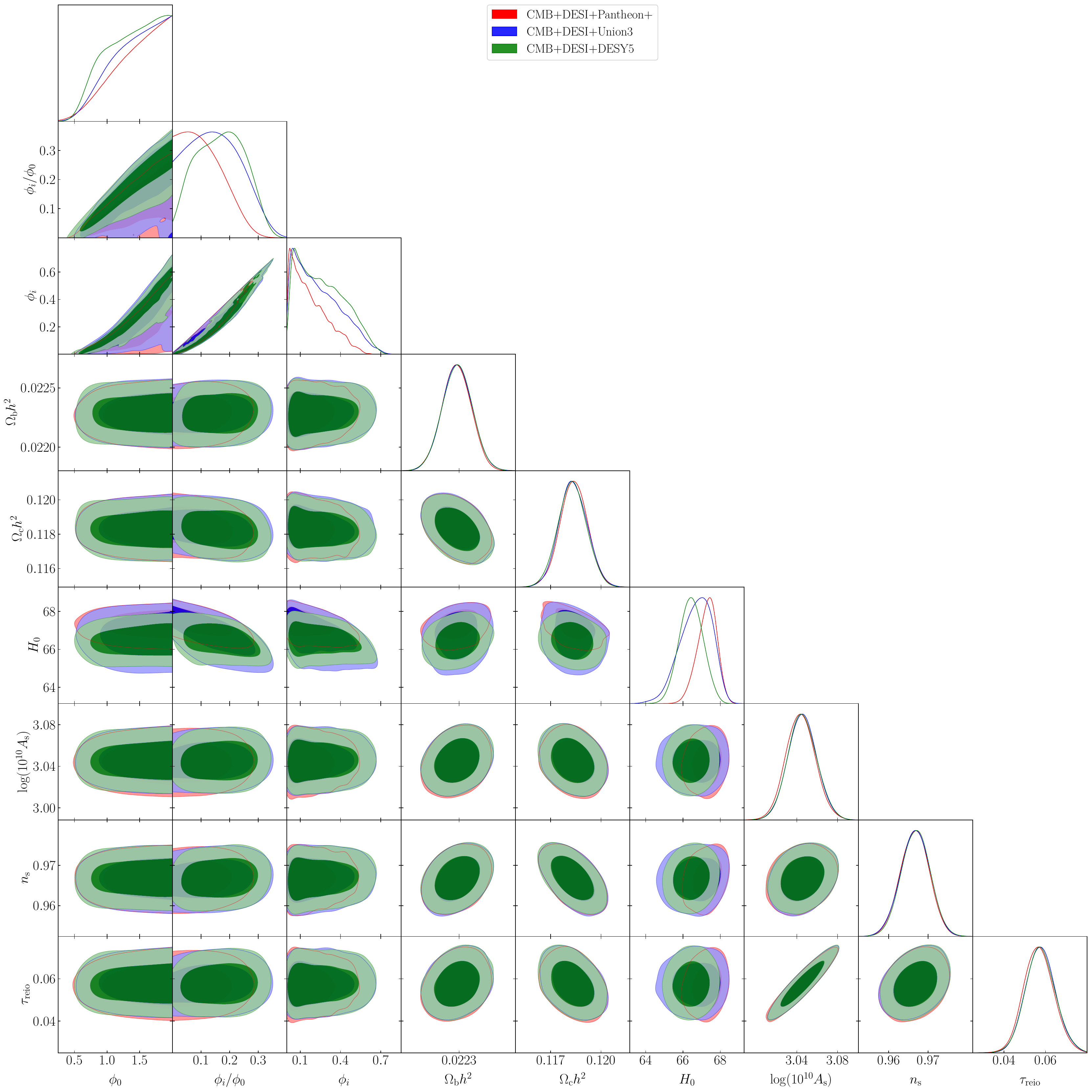}
    \caption{Full parameter constraints for the hilltop quartic model.}
    \label{fig:Higgs_mcmc}
\end{figure}

\begin{table}[]
    \centering
\begin{tabular} {|c| c| c| c|}

\hline
\rowcolor{gray!30} 
 \bf{Parameter} &  \textbf{+Pantheon+} &  \textbf{+Union3} &  \textbf{+DESY5}\\
\hline
{\boldmath$\Lambda         $} & $> 1.29    \ml{1.23}                $ & $> 1.24  \ml{0.69}                  $ & $> 1.17  \ml{0.62}                  $\\

{\boldmath$\phi_i/\Lambda  $} & $< 0.142  \ml{0.106}                 $ & $0.151^{+0.073}_{-0.12}  \ml{0.051}  $ & $0.169\pm 0.081  \ml{0.034}          $\\

{\boldmath$\Omega_\mathrm{c} h^2$} & $0.11838\pm 0.00082 \ml{0.1188}       $ & $0.11835\pm 0.00084   \ml{0.1176}     $ & $0.11829\pm 0.00084  \ml{0.1189}      $\\

{\boldmath$\Omega_\mathrm{b} h^2$} & $0.02228\pm 0.00012   \ml{0.02221}     $ & $0.02228\pm 0.00013 \ml{0.00223}       $ & $0.02228\pm 0.00013  \ml{0.02223}      $\\

{\boldmath$\log(10^{10} A_\mathrm{s})$} & $3.044\pm 0.014 \ml{3.041}           $ & $3.046\pm 0.014     \ml{3.044}       $ & $3.045\pm 0.014    \ml{3.033}        $\\

{\boldmath$n_\mathrm{s}   $} & $0.9666\pm 0.0036   \ml{0.9641}      $ & $0.9667\pm 0.0037  \ml{0.9694}        $ & $0.9669\pm 0.0036  \ml{0.9645}         $\\

{\boldmath$H_0            $} & $67.29^{+0.59}_{-0.45} \ml{66.98}    $ & $66.7^{+1.0}_{-0.70}     \ml{66.30}  $ & $66.44\pm 0.64   \ml{66.06}          $\\

{\boldmath$\tau_\mathrm{reio}$} & $0.0567\pm 0.0071  \ml{0.0555}        $ & $0.0576\pm 0.0071       \ml{0.056}   $ & $0.0576\pm 0.0072  \ml{0.0514}        $\\

$\phi_i                  $ & $0.174^{+0.071}_{-0.17} \ml{0.1313}     $ & $0.235^{+0.088}_{-0.23}  \ml{0.0035}  $ & $0.26^{+0.10}_{-0.24}  \ml{0.0021}    $\\
\hline
\end{tabular}
    \caption{Hilltop quartic model: full parameter means and $68\%$ limits for the addition of the different supernovae datasets to the CMB+DESI combination. The values in parentheses denote the best-fit parameters for this model.}
    \label{tab:Higgs_table_full}
\end{table}

\clearpage

\subsection*{Hilltop octic potential}

\begin{figure}[H]
    \centering
    \includegraphics[width=0.95\linewidth]
    {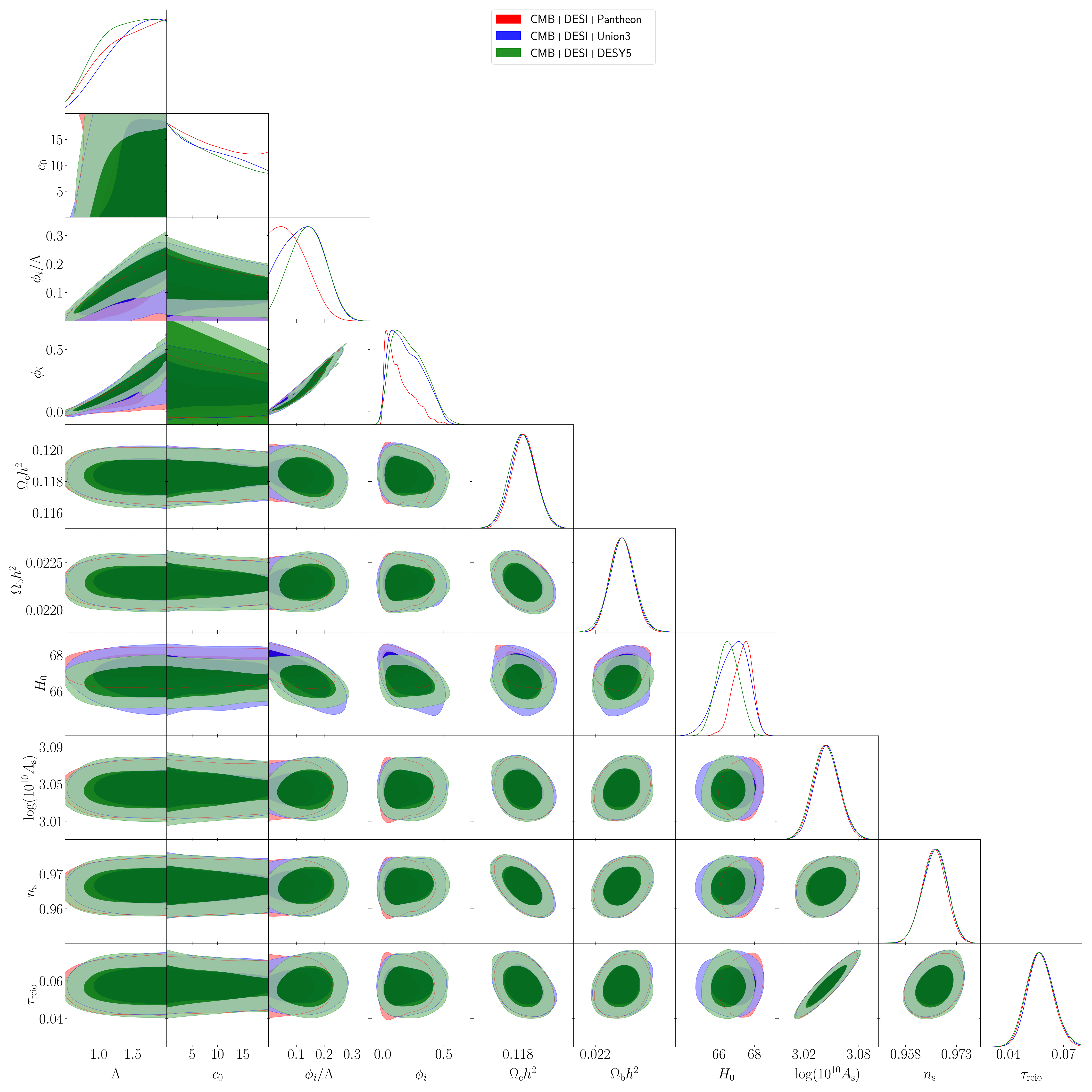}
    \caption{The complete parameter constraints for the hilltop octic model.}
   \label{Exp_triangle_full}
\end{figure}

\begin{table}[]
    \centering
\begin{tabular} {|c| c| c| c|}

\hline
\rowcolor{gray!30} 
 \bf{Parameter} &  \textbf{+Pantheon+} &  \textbf{+Union3} &  \textbf{+DESY5}\\
\hline
{\boldmath$\Lambda              $} & $> 1.21  \ml{0.995}                 $ &  $> 1.29  \ml{1.195}                $ & $> 1.18 \ml{0.826}      $\\

{\boldmath$c_0              $} & ---  $\ml{8.91}                 $ & $< 12.4  \ml{14.27}                $ & $< 11.8 \ml{9.78}      $\\

{\boldmath$\phi_i/\Lambda     $} & $< 0.113      \ml{0.067}                $    & $0.127\pm 0.067     \ml{0.133}                 $    & $0.141\pm 0.062    \ml{0.071}                $\\

{\boldmath$\Omega_\mathrm{c} h^2$} & $0.11839\pm 0.00079  \ml{0.1181}      $ & $0.11835\pm 0.00084  \ml{0.1182}      $ & $0.11828\pm 0.00084 \ml{0.1178}       $\\

{\boldmath$\log(10^{10} A_\mathrm{s})$} & $3.044\pm 0.014  \ml{3.048}          $ & $3.045\pm 0.014 \ml{3.049}           $ & $3.044\pm 0.015 \ml{3.048}           $\\

{\boldmath$n_\mathrm{s}   $} & $0.9664\pm 0.0035  \ml{0.967}       $ & $0.9666\pm 0.0036  \ml{0.967}        $ & $0.9666\pm 0.0037   \ml{0.967}       $\\

{\boldmath$H_0            $} & $67.34^{+0.59}_{-0.47}  \ml{67.31}   $ & $66.78^{+0.97}_{-0.70} \ml{65.47}    $ & $66.50\pm 0.62 \ml{65.87}    $\\

{\boldmath$\Omega_\mathrm{b} h^2$} & $0.02229\pm 0.00012  \ml{0.02229}      $ & $0.02229\pm 0.00013 \ml{0.02225}      $ & $ 0.02229\pm 0.00014 \ml{0.02228}       $\\

{\boldmath$\tau_\mathrm{reio}$} & $0.0571\pm 0.0069 \ml{0.0581}$ & $0.0577\pm 0.0069 \ml{0.0589}$ & $0.0571^{+0.0069}_{-0.0078} \ml{0.0595}         $\\

$\phi_i                  $ & $0.136^{+0.044}_{-0.14}  \ml{0.067}      $ & $0.199^{+0.091}_{-0.17} \ml{0.016}      $ & $0.214^{+0.097}_{-0.17} \ml{0.059}       $\\

\hline
\end{tabular}
    \caption{Hilltop octic model: full parameter means and $68\%$ limits for the addition of the different supernovae datasets to the CMB+DESI combination. The values in parentheses denote the best-fit parameters for this model.}
    \label{tab:Oct_table_full}
\end{table}
\clearpage

\subsection*{Hilltop exponential potential}
\begin{figure}[H]
    \centering
    \includegraphics[width=0.95\linewidth]
    {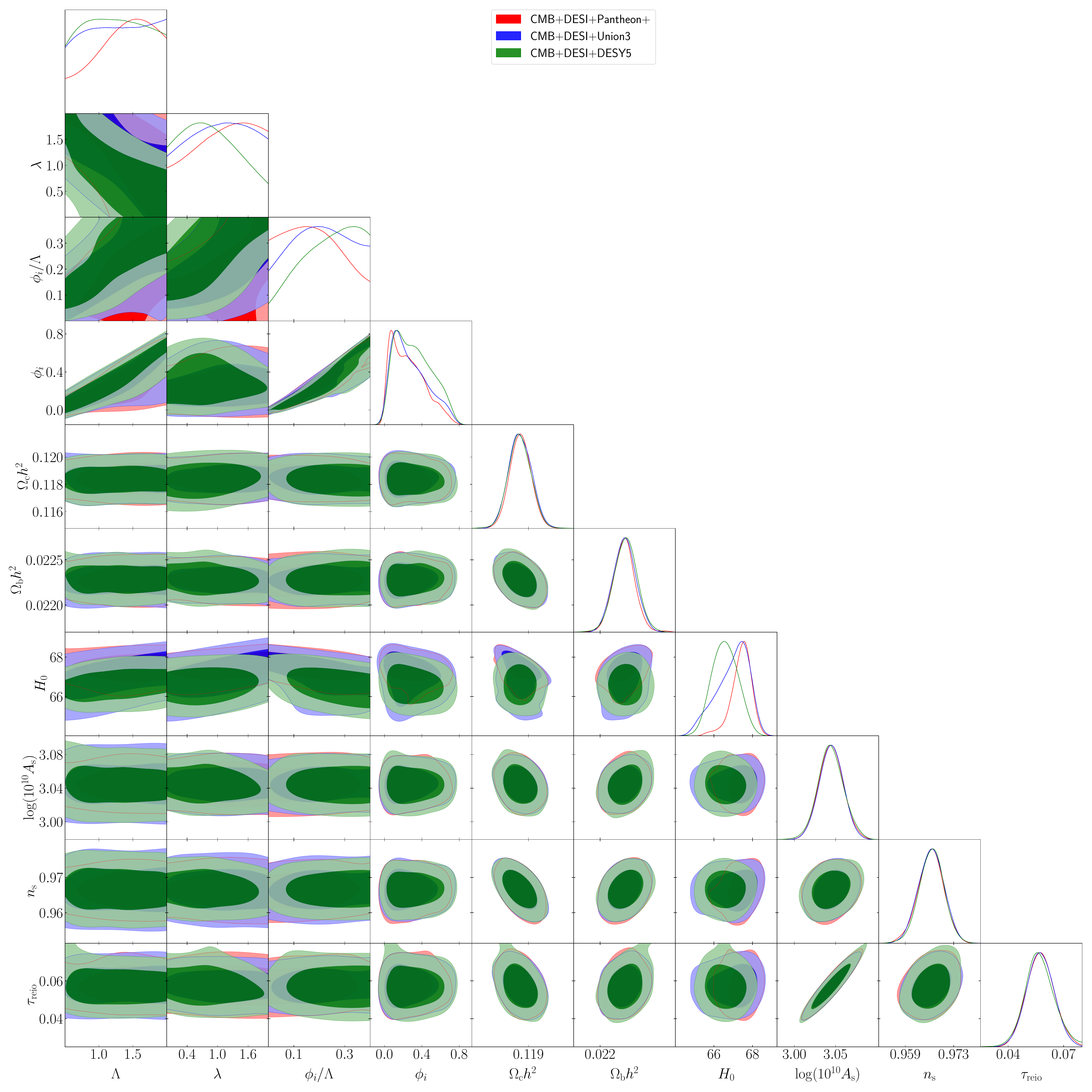}
    \caption{The complete parameter constraints for the hilltop exponential model.}
   \label{Exp_triangle_full}
\end{figure}

\begin{table}[]
    \centering
\begin{tabular} {|c| c| c| c|}

\hline
\rowcolor{gray!30} 
 \bf{Parameter} &  \textbf{+Pantheon+} &  \textbf{+Union3} &  \textbf{+DESY5}\\
\hline
{\boldmath$\Lambda              $} & $1.36^{+0.64}_{-0.20}  \ml{0.595}                 $ &  --- $ \ml{0.742}                $ & --- $ \ml{0.524}      $\\

{\boldmath$\lambda              $} & $> 0.823  \ml{1.77}                 $ & --- $ \ml{1.76}                $ & $0.89^{+0.36}_{-0.80} \ml{1.95}      $\\

{\boldmath$\phi_i/\Lambda     $} & $0.180^{+0.068}_{-0.17}      \ml{ 0.243}                $    & $0.21\pm 0.11     \ml{0.333}                 $    & $> 0.183    \ml{ 0.301}                $\\

{\boldmath$\Omega_\mathrm{c} h^2$} & $0.11840\pm 0.00079 \ml{0.1179}      $ & $0.11839\pm 0.00085  \ml{0.1179}      $ & $0.11835\pm 0.00085 \ml{0.1184}       $\\

{\boldmath$\log(10^{10} A_\mathrm{s})$} & $3.045\pm 0.015 \ml{3.050}          $ & $3.045\pm 0.014 \ml{3.046}           $ & $3.044\pm 0.016 \ml{3.037}           $\\

{\boldmath$n_\mathrm{s}   $} & $00.9664\pm 0.0037  \ml{0.968}       $ & $0.9665\pm 0.0037  \ml{ 0.966}        $ & $0.9666\pm 0.0036   \ml{0.967}       $\\

{\boldmath$H_0            $} & $67.42^{+0.57}_{-0.39}  \ml{67.31}   $ & $67.0^{+1.1}_{-0.60} \ml{66.04}    $ & $66.58\pm 0.67 \ml{66.43}    $\\

{\boldmath$\Omega_\mathrm{b} h^2$} & $0.02227\pm 0.00013  \ml{0.02233}      $ & $0.02228\pm 0.00013 \ml{0.02234}      $ & $ 0.02229\pm 0.00013 \ml{0.02232}       $\\

{\boldmath$\tau_\mathrm{reio}$} & $0.0574\pm 0.0075 \ml{0.0607}$ & $0.0573\pm 0.0073 \ml{0.057}$ & $0.0571^{+0.0073}_{-0.0083} \ml{0.0522}         $\\

$\phi_i                  $ & $0.25^{+0.10}_{-0.23}  \ml{0.145}      $ & $0.28^{+0.12}_{-0.24} \ml{0.247}      $ & $0.31^{+0.14}_{-0.26} \ml{0.158}       $\\

\hline
\end{tabular}
    \caption{Hilltop exponential model: full parameter means and $68\%$ limits for the addition of the different supernovae datasets to the CMB+DESI combination. The values in parentheses denote the best-fit parameters for this model.}
    \label{tab:Exp_table_full}
\end{table}

\clearpage

\subsection*{PySR potential}
{In analyzing  this model we only explore the parameter space for the CMB+DESI+Union3 combination. This is because the SR was performed assuming the DESI reconstruction of $w(z)$, which is based on the same data combination. Therefore, applying the same analysis to the CMB+DESI+Pantheon+ and CMB+DESI+DESY5 combinations, as we did for the other models, would be inconclusive.}
\begin{figure}[H]
    \centering
    \includegraphics[width=0.95\linewidth]
    {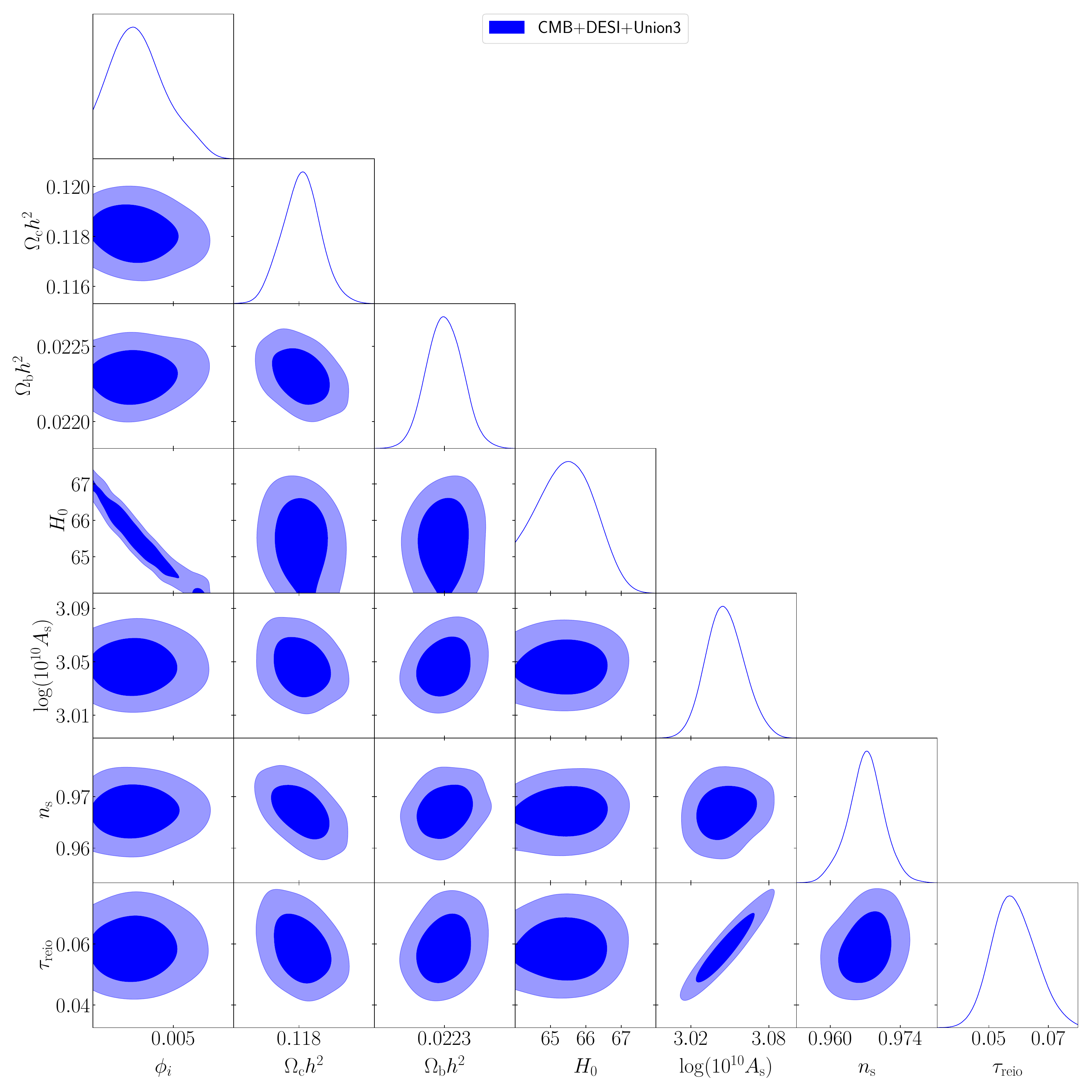}
    \caption{The parameter constraints for the PySR model.}
   \label{PySR_triangle_full}
\end{figure}

\begin{table}[H]
    \centering
\begin{tabular} {|c| c|}

\hline
\rowcolor{gray!30} 
 \bf{Parameter}  &  \textbf{+Union3}\\
\hline

{\boldmath$\Omega_\mathrm{c} h^2$} & $0.11810\pm 0.00078 \ml{0.1182}$\\

{\boldmath$\log(10^{10} A_\mathrm{s})$} & $3.046^{+0.014}_{-0.016}   \ml{3.042}   $\\

{\boldmath$n_\mathrm{s}   $} & $0.9671\pm 0.0036 \ml{0.966} $\\

{\boldmath$H_0            $} & $65.48\pm 0.75 \ml{65.04}$\\

{\boldmath$\Omega_\mathrm{b} h^2$} & $ 0.02230\pm 0.00012 \ml{0.02228} $\\

{\boldmath$\tau_\mathrm{reio}$} &  $0.0585^{+0.0068}_{-0.0079}    \ml{0.0567}     $\\

$\phi_i                 $ & $0.00377^{+0.00072}_{-0.0013}  \ml{0.0042}$\\

\hline
\end{tabular}
    \caption{PySR model: parameter means and $68\%$ limits for the CMB+DESI+Union3 combination. The values in parentheses denote the best-fit parameters for this model.}
    \label{tab:PySR_table_full}
\end{table}

\end{appendix}

{\small
\addcontentsline{toc}{section}{ References}
\bibliographystyle{utphys}
\bibliography{boundedDE}
}

\end{document}